\documentstyle[12pt,aaspp]{article}

\def\etal{{\it et~al.~}}
\def\eg{{\it e.g.~}}
\def\ie{{\it i.e.~}}

\begin{document}

\title{Globular Clusters with Tidal Tails: Deep Two-Color Star Counts}
\author{Carl J. Grillmair\altaffilmark{1}, K. C. Freeman}
\affil{Mount Stromlo and Siding Springs Observatories, Australian
National University, Private Bag, Weston Creek Post Office, ACT 2611,
Australia}

\altaffiltext{1}{Current Address: Lick Observatory,University of
California, Santa Cruz, CA 95064}

\author{M. Irwin}
\affil{Institute for Astronomy, Madingley Road, Cambridge, CB3 0EZ,
United Kingdom}

\author{P. J. Quinn}
\affil{Mount Stromlo and Siding Springs Observatories, Australian
National University, Private Bag, Weston Creek Post Office,
ACT 2611, Australia}

\begin{abstract}

	We examine the outer structure of 12 Galactic globular
clusters using star-count analyses. Deep, two-color, photographic
photometry is used to select and count stars with colors and
magnitudes consistent with cluster-specific, color-magnitude
sequences. The resulting reduction in the number of contaminating
foreground stars allows us to push the star counts to significantly
lower surface densities than has previously been possible. We find
that most of our sample clusters show extra-tidal wings in their
surface density profiles. The form of the surface density profiles is
consistent with recent numerical studies of the tidal stripping of
globular clusters. Two-dimensional surface density maps for several
clusters are consistent with the expected appearance of tidal tails,
with allowance for the effects of orbit shape, orbital phase, and
orientation of our line of sight.  We identify the extra-tidal
material with stars which are still in the process of being removed
from the clusters. The extra-tidal stars effectively limit the
accuracy to which we can resolve the ``tidal'' radii of globular
clusters. However, by tracing the orbital paths of their parent
clusters, these stars may also facilitate a renewed attack on the
problem of determining globular cluster orbits and the shape of the
Galactic potential field.

\end{abstract}


\section{Introduction.}

	Globular clusters first demonstrated their usefulness as
tracers of Galactic structure when Harlow Shapley located the center
of the Galaxy by studying their distribution in space (Shapley 1918).
Since that time, globular clusters have in many ways become the
cornerstone for our understanding of the formation, structure, and
dynamics of the halo of our Galaxy. We were initially motivated to
undertake a study of the limiting radii of globular clusters by the
curious finding of Seitzer (1983) (see also Freeman and Norris 1981)
that the metallicities of clusters appeared to be correlated with
their orbital radii at perigalacticon, $R_{min}$. Clusters whose
orbits are believed to take them nearer to the Galactic center
generally have higher metallicities.  Such a correlation is not nearly
as evident in a plot of metallicity against present Galactocentric
distance, and one possible inference is that clusters may have
actually formed near their perigalactica (Freeman and Norris 1981).
Since the tidal radii of globular clusters depend in some manner on
the perigalactic distances of their orbits, our hope was to improve on
existing determinations of $R_{min}$ and hence to both verify and
examine more closely this relationship with metallicity.

 It has long been supposed that globular clusters must have a finite
edge due to the removal of stars by the Galactic tidal field. However,
establishing the radius at which cluster surface densities actually
vanish has proven difficult.  Owing to the remarkable agreement over
several orders of magnitude between the projected surface densities
predicted by King models (King 1966) and most observed cluster surface
density profiles, limiting radii of globular clusters have
traditionally been estimated using the model-predicted value of the
King tidal radius, $r_{t}$. Based on the postulates that globular
clusters are limited in extent by the Galactic tidal field and that
the limiting radii are established at the perigalactic points of
cluster orbits (von Hoerner 1957; King 1962), Peterson (1974)
attempted to put constraints on the shapes of globular cluster orbits
using the then published values of $r_t$. The results of this work, while
suggestive, were not entirely credible since the inferred perigalactic
distances for several clusters were found to be greater than their
present Galactocentric distances. Innanen, Harris, \& Webbink (1983)
carried out a similar study and concluded that globular cluster orbits
were rather more circular than an isotropic velocity distribution
would require.  However, their attempts to determine individual
cluster perigalactica were defeated, and they cited the large
uncertainties in published values of $r_t$ as being chiefly responsible.

 	In this paper we use the method of star-counts to examine
the peripheral structure of a sample of Galactic globular clusters.
Star-count analyses are statistically superior to aperture photometry
at large radii where the ratio of cluster surface brightness to sky
brightness becomes smaller than the ratio of cluster stars to
foreground stars. Moreover, ignoring noise contributions from the
foreground stars, the star-count signal-to-noise ratio goes as
$\sqrt{N}$, where $N$ is finite, while the noise associated with
aperture photometry can be heavily influenced by relatively few
luminous giant stars.

   	 Assuming for the moment that $r_t$ should actually correspond
to the real, physical limits of clusters, its fitted value is heavily
influenced by the superior statistics available at relatively small
radii. The literature contains several examples in which the star
counts near $r_{t}$ apparently depart from the form predicted by the
King model which best fits the bulk of the data (Peterson 1976;
Illingworth \& Illingworth 1976).  Indeed, Peterson \& King (1975)
note several instances in which King models do not fit the star-count
profiles at all. In view of the relatively minor consequences for
studies of internal cluster dynamics, these departures have never been
seriously followed up.

	The largest source of uncertainty in attempting to measure
surface density profiles near the limiting radii of globular clusters
results from simple Poisson statistics introduced by the overwhelming
number of contaminating field stars. An obvious way to reduce the
number of field stars in a given sample is to count only stars which
have colors and magnitudes which resemble those of cluster stars.
Prior to embarking on this project, we carried out extensive
simulations to determine the extent to which tidal radius
determinations could be improved by being so selective. Using King
model surface density distributions, canonical cluster color-magnitude
sequences, and field star distributions computed from the
Bahcall-Soneira model (Bahcall \& Soneira 1980; Mamon \& Soneira
1982), the simulations revealed that a large fraction of the field
stars could indeed be eliminated from the counting process. Combining
color-selection with a more direct method of fitting the surface
density profile at large radii, we found that the uncertainty in the
fitted value of $r_{t}$ could be reduced by as much as an order of
magnitude.  The working assumption in these simulations was that the
cluster surface density profiles at large radii could be well
represented by King models...

	The observational material used for this study is described in
Section \ref{sec:obs}. Star counts, residual foreground removal, and
crowding corrections are discussed in Section \ref{sec:starcounts}.
In Section \ref{sec:tidalradii} we examine the observed structure of
globular clusters at large radii. We briefly compare our observational
findings with recent results of numerical simulations in Section
\ref{sec:discussion}.

\section{Observations. \label{sec:obs} }

	The sample we have chosen to study consists of 12 Galactic
globular clusters with current Galactocentric distances in the range 9
$< R_{GC} < 40$ kpc.  These clusters are particularly interesting in
that their metallicities appear to be related to their perigalactic
distances, as inferred from their fitted tidal radii (Freeman and
Norris 1981). Table 1 lists relevant data from the literature
concerning these clusters.  Coordinates are taken from Shawl \& White
(1986). Core and tidal radii are from the compilation of Trager,
Djorgovski, \& King (1993), and values for $R_{GC}$ are taken from the
compilation of Djorgovski (1993).

	Our simulations revealed that, in most cases, the accuracy to
which we could resolve the tidal cutoffs of model clusters depended
upon pushing the star counts as far down the main sequence as
possible.  This is purely a consequence of the large rise in the
luminosity function at faint magnitudes and the correspondingly
improved counting statistics.  Exceptions to this general rule were
dictated by the predicted onset of excessive numbers of field stars
within some envelope containing the cluster-specific color-magnitude
sequence.  These exceptions included clusters situated (in projection)
near the Galactic plane (NGC 2808, NGC 3201) or the Small Magellanic
Cloud (NGC 362).

	Despite the advent of large-format CCDs, the apparent sizes of
globular clusters are so large that the only practical way of carrying
out a study of this type is through the use of the photographic
Schmidt plate. Our desire to go as deep as possible, combined with the
fact that none of the sample clusters is well-centered on existing
Survey plates, prompted us to apply for suitable plate material to be
taken with the United Kingdom Schmidt Telescope (UKST) at Siding
Springs Observatory. Thanks to the dedicated efforts of the staff of
the UKST, some 50 high-quality plates were acquired in the space of
just over a year. The details concerning these plates are given in
Table 2.  Whereas most cluster plates were sky-limited, the
exceptional fields noted above required exposure times limited to
about 30 minutes.  By request, the plates were typically taken in
conditions during which the seeing FWHM was $>$ 2\arcsec. In addition
to improving the photometric accuracy, this enabled us to take
advantage of periods during which seeing conditions were inadequate
for higher-priority Southern Survey plates.  On the other hand, it
also meant that crowding of stellar images near the cores of the
clusters would be correspondingly more severe. Two plates were
acquired in each of $B_J$ and $R$ to improve the photometry at faint
magnitudes.

 The plates were scanned using the the Automatic Plate Measuring
System (APM, (Kibblewhite \etal 1984). This system  has been in
operation for over a decade and has become an invaluable resource for
large-scale photographic survey work. Its primary features include a
laser spot scanner capable of scanning an entire Schmidt plate in
about 2 hours, as well as extensive on-line processing capabilities.
The actual scanning and processing procedures have been discussed at
length by Irwin \& Trimble (1984) and Bunclark \& Irwin (1983) and are
detailed here only as they become relevant.

	Processing of the plate scans was carried out using a
crowded-field algorithm (Irwin 1985) so as to push the star counts as
far into the cores of the clusters as possible.  Based on areal
profiles and second order moments, detected images were classified as
either stars, merged images, non-stellar sources, or noise. For each
plate, scanning was carried out for an area about $4^{\circ}$ square
and centered on the cluster. The total number of detected images
ranged from 60,000 to 400,000 per field.

\section{Star Counts. \label{sec:starcounts} }

\subsection{Identification of Cluster Stars. \label{sec:ident} }

	In Figure \ref{fig:cofm} we show color-magnitude diagrams,
based on the APM magnitudes of images classified as stars, both for
the central regions of each cluster and for annuli extending outwards
from well beyond the published tidal radii. The APM magnitudes are
linearly related to real apparent magnitudes over essentially the
entire range of magnitudes shown in these figures (Bunclark \& Irwin
1983). Where we have two or more magnitudes available per image in
each color, we have used the average value.  Images for which we have
no color information (due, for example, to differences in image size
or background density between $B_J$ and $R$ plates) are not used. The
cluster sequences, particularly near the turnoff region, are quite
distinct. Based on the width of the main sequences, we estimate our
magnitude uncertainties to be about 0.2 mag at the limit of the
deepest plates. The zero-point calibration is approximate but is
essentially irrelevant for the purposes of establishing cluster
membership.

	Shown as heavy lines in Figure \ref{fig:cofm} are the regions
within which stars are considered to be likely cluster members for
counting purposes. Whereas the color-magnitude sequences evident in
the left-hand panels of Figure \ref{fig:cofm} rely on a few of the
most crowded and least well-measured stars on each plate, defining the
color-magnitude envelope is carried out for a much larger number of
stars over a much larger area of sky. The color-magnitude envelopes
are empirically chosen so as to optimize the ratio of cluster stars to
field stars in the relatively sparsely populated outer regions of each
cluster.  This is done by subdividing the
color-magnitude plane into a 50$\times$50 array in which individual
elements are 0.08 magnitudes wide in color and 0.2 magnitudes high in
APM $B_J$.  Assuming that the color-magnitude distribution of the
field stars does not vary across the plate, an APM color-magnitude
sequence for each cluster can be estimated from

\begin{equation} f_{cl}(i,j) = n_{cl}(i,j) - gn_{f}(i,j)
\label{eq:cm}
\end{equation}

\noindent where $n_{cl}, n_{f}$ refer to the number of images with
color index $i$ and magnitude index $j$ counted within the central
region of the cluster and in an annulus well outside the cluster,
respectively. $g$ is simply the ratio of the area of the cluster
annulus to that of the field-star annulus. We compute the
``signal-to-noise'' ratio for each color-magnitude element

\begin{equation} s(i,j) = \frac{f_{cl}(i,j)}{\sqrt{n_{cl}(i,j) +
g^{2}n_{f}(i,j)}}.
\label{eq:sna}
\end{equation}

\noindent The $s(i,j)$ array is then smoothed according to the
estimated color and magnitude errors in each element. Errors in $B_J$
and $R$ as a function of magnitude are estimated by comparing the
magnitudes determined from different plates of the same color. Each
element of the color-magnitude array is then convolved with a
bivariate, Gaussian smoothing kernel with dispersions in the $B_J$ and
$B_J - R$ dimensions set to be equal to the appropriate errors. Figure
\ref{fig:n3201cmmap} shows a contour map of the smoothed distribution
of $f_{cl}(i,j)$ for NGC 3201. Shown as a heavy line is a contour of
constant signal-to-noise ratio, $s(i,j)$ (smoothed), arbitrarily chosen
to match the contours of $f_{cl}(i,j)$ on the red side of the cluster
sequence. The slight blueward bias of $s(i,j)$ is a consequence of the
relatively low density of field stars blueward of the cluster sequence
(\eg Figure \ref{fig:cofm}), and the presence of disproportionate
numbers of field stars to the red.

	The optimal range of colors and magnitudes for each cluster
were determined as follows. ($i$) The elements of $s(i,j)$ were sorted
into descending order over the one-dimensional index $l$.  ($ii$)
Beginning with the array element with the highest signal-to-noise
ratio ($l$=1), star counts were carried out in an test annulus
extending over $r_{t}/2 \leq r \leq r_{t}$ using progressively larger
areas of the color-magnitude diagram $a_{k} = ka_{l}$, where
$a_{l} = 0.016$ mag$^{2}$ is the area of a single element in the
color-magnitude array. ($iii$) Field-star surface densities were
determined as a function of $a_{k}$ in the same manner. ($iv$) The
cumulative signal-to-noise ratio, $S(a_{k})$, was computed from

\begin{equation} S(a_{k}) = \frac{N_{cl}(a_{k}) - gN_{f}
(a_{k})}{\sqrt{N_{cl} (a_{k})  + g^{2} N_{f} (a_{k})}}
\label{eq:snr}
\end{equation}

\noindent where

\begin{equation}
 N_{cl}(a_{k}) = \sum_{l=1}^{k} n_{cl}(l)
\label{eq:snra} \end{equation}

\begin{equation} N_{f}(a_{j}) = \sum_{l=1}^{k} n_{f}(l)
\label{eq:snrb} \end{equation}

\noindent and $n_{cl}(l)$ now refers to the number of images within
the test annulus having ordered color-magnitude index $l$.  In
principle one could normalize the distribution of stars near the
cluster center to the surface density of cluster stars in the outer
regions (rather than counting stars in a test annulus) to achieve a
statistically smoother result, but one would then have to contend with
crowding and luminosity effects which are themselves functions of
surface density.  In either case, the lower cluster-star surface
density has the effect of moving the peak in the $S(a_k)$ function to
smaller color-magnitude envelopes.  Figure \ref{fig:n362snvsarea}
shows the run of $S$ with increasing color-magnitude area $a_{k}$ for
the case of NGC 362. As is apparent in Figure \ref{fig:cofm}, the
color-magnitude sequence of NGC 362 is hemmed in by foreground
Galactic stars and the giant branch and blue main sequence of the
Small Magellanic Cloud. $S$ consequently drops quite quickly as one
moves laterally away from the cluster sequence. ($v$) Based on the
peak value of $S$ as read from plots such as Figure
\ref{fig:n362snvsarea}, a threshold value of $s \ge s(i,j)$ is
determined, yielding one or more regions in the color-magnitude array
in which the contrast between the surface densities of cluster and
field stars is highest.  ($vi$) Minor editing of the perimeter of the
optimal color-magnitude envelope was carried out to remove outliers
({\it e.g.} single-star events in areas of the color-magnitude diagram
where the field-star density is low) and to smooth the contours of the
selected area.

In several instances, the color-magnitude envelopes ultimately
selected included only stars at and below the turnoff and,
occasionally, on the horizontal branch.  This reflects both the rise
in the luminosity function at faint magnitudes and the fact that the
cluster giant branches often extend into regions of the
color-magnitude diagram heavily populated by disk main-sequence
stars. Blue horizontal branches fall well clear of the main body of
foreground stars and thus provide an inexpensive additional source of
counts. While some of the color-magnitude sequences in Figure
\ref{fig:cofm} may appear somewhat ratty due to high field-star
contamination, severe image crowding, and plate saturation, the
selected color-magnitude envelopes do not rely on these sequences and
are considerably more secure. Our final color-magnitude envelopes
reduce foreground contamination by between 41\% (NGC 3201) and 92\%
(NGC 7078).

\subsection{Modeling the Distribution of Field Stars.
\label{sec:foreground} }

\indent	Each $\approx4^{\circ}$ square scan was subdivided into an array
of 128$\times$128 elements. Surface densities were computed by simply
counting the number of images in each element and dividing by the
appropriate area. All images (rather than just those classified as
stars) were counted owing to the rather unpredictable performance of
the image classification algorithm in regions of high surface density.
In regions of low surface density, the inclusion of images classified
as non-stellar contributed to the Poisson uncertainties in
the counts. However the mean surface densities of non-stellar images
in regions beyond the published tidal radii never constituted more than
25\% of the total so their contribution to the counting statistics is
relatively minor.  The contribution of galaxies to the variation in
surface density across the field is considered in Section
\ref{sec:2dst}

 Foreground contamination was modeled by masking the central region
of each plate scan (generally the region within $1.5r_{t}$) and fitting
a low-order, bivariate polynomial to the remaining area.  This
polynomial has the form

\begin{equation}
z(x,y) = \sum_{i=0}^{k} \sum_{j=0}^{l} a_{ij}x^{i} y^{j}
 \label{eq:bipoly}
\end{equation}

\noindent where $k$ and $l$ refer to the degree of the polynomial in the
$x$ and $y$ directions, respectively.  Thus, a 1$\times$1 polynomial
fit would be a twisted plane in which the $z(x,y)$ varies linearly along
any given row or column.

	The star counts near the limiting radii of globular clusters
are sensitive to the form of foreground subtraction used.
Historically, star-counts have been foreground-subtracted using the
surface density, integrated around an annulus, at a radius deemed
large. In the absence of foreground density fluctuations on scales
similar to that of the cluster itself, and assuming that the
determination of the foreground density is indeed carried out well
beyond the limits of the cluster, this method should in principle be
perfectly adequate. However, as we shall see in Section
\ref{sec:tidalradii}, variations in the distribution of foreground
stars can be quite dramatic, whether due to differential extinction
near the galactic plane or significantly non-linear surface density
gradients across the field. An extreme example is that of NGC 362,
which is projected against the northern extension of the Small
Magellanic Cloud.

  The order of the surface fits were chosen so that the resulting sky
arrays are dominated by the smooth gradient in field density due to
underlying structure of the Galaxy. In Figure \ref{fig:n7089sky} we
compare the model foreground distribution computed for the field
surrounding NGC 7089 using 1$\times$1 and 3$\times$3 polynomials. The
cross terms in Equation \ref{eq:bipoly} consequently admit variations
of 2nd and 6th order, respectively. While the 1st-order fit reflects
the gradient expected from the Galactic distribution, a 3rd-order fit
is heavily influenced by local concentrations which may or may not be
related to the cluster itself (see below). In all but three cases, a
1$\times$1 degree polynomial fit was sufficient to remove any
large-scale gradients.  Fields near the Galactic plane are fitted with
2$\times$2 bivariate polynomials owing to the underlying exponential
increase in field-star densities across the plate.  For the field
surrounding NGC 362 we use a 2$\times$3 bivariate polynomial fit to
model the surface density distribution of the northern wing of the
SMC.  In a few instances, small regions of the plate scan containing
obvious concentrations of images were masked to avoid skewing the
surface fit.  With one exception (NGC 7078) the maximum gradients in
these surface fits correspond well with the direction to the galactic
plane.

	Star counts were carried out using a software
r\'{e}seau with concentric annuli one arcminute wide and centered on
the cluster coordinates.  Since the central 2\arcmin~ of our
clusters are saturated or crowded beyond the ability of the APM
software to distinguish individual images, the coordinates of the
cluster centers have been taken from Shawl \& White (1986).  The raw
star counts, along with computed surface densities, associated
uncertainties, and appropriate crowding corrections (see Section
\ref{sec:crowding}) are given in Tables 3 through
14. The table entries are arranged as follows: The
first two columns specify the inner and outer radii of individual
annuli in arcminutes. Beyond 9\arcmin~ the annuli have been combined
into progressively larger bins to reduce the relative uncertainties
for individual entries. In column 3 are given the effective radii
appropriate to each entry. The tabulated radii are area-weighted
rather than luminosity weighted ({\it e.g.} King 1988), but the
surface density gradients over the range of radii with which we are
concerned are sufficiently small to make such a distinction
insignificant.  Column 4 contains the actual number of images counted.
In column 5 we tabulate the mean field-star densities determined from
surface fitting and corrected for crowding.  Crowding corrections,
which are given as surface density multipliers, are tabulated in
column 6. Note that whereas the human eye is rarely confused by image
blending in regions of low surface density and that crowding
corrections would not generally be required, machine counts require
correction at all radii. We discuss crowding corrections at length in
Section \ref{sec:crowding}. The sky-subtracted surface densities,
corrected for crowding, are given along with their corresponding
uncertainties (computed purely from Poisson statistics) in columns 7
and 8.

	At the end of each table are indicated the sectors used for
the star counts. The r\'{e}seau used for counting is divided into
eight sectors with opening angles of 45$^{\circ}$, as shown in Figure
\ref{fig:sectors}. In a few instances, certain sectors were avoided
owing to either obvious concentrations of images unrelated to the
cluster in question or uncertainties in the foreground surface fitting
of specific regions. Also indicated at the bottom of each table are
the degree of the surface-fitting polynomial used and the average
image area $A$ used to compute the crowding corrections.

The tabulated surface densities are shown plotted in Figure
\ref{fig:sd} along with normalized star-count profiles of King \etal
(1968) and Peterson (1976), and surface densities computed from
Peterson's (1986a) compilation of photoelectric aperture photometry.
We plot our data only out to the radius beyond which the computed
surface densities are first found to be less than one error bar above
zero.  Where King \etal provide more than one star-count profile, we
have plotted the two profiles with the highest fitted background
densities and/or largest radial extents.  The aperture photometry data
are plotted at their King luminosity-weighted radii.  Also shown in
these figures are King models computed from the published core and
tidal radii (see Table 1), arbitrarily normalized to our counts near
the limiting radius. Data sets from the literature are independently
normalized to these model-predicted surface densities for all clusters
except NGC 5824. Due to the mismatch between the observed and
predicted profiles, the aperture photometry for NGC 5824 are instead
normalized so as to match the star counts in the region of
overlap. References to the original sources of data are given in Table
15.

\subsection{Crowding Corrections. \label{sec:crowding} }

	Owing to the marginal seeing conditions under which our
cluster plates were taken, we are rather severely affected by image
crowding at high surface densities. This is evident in Figure
\ref{fig:sd}, where the $\times$'s denote the raw star counts
uncorrected for the effects of image blending. The filled circles
result from applying the crowding correction formula derived for
machine counts by Irwin \& Trimble (1984)

\begin{equation}
\alpha_{c} = \frac{f'}{f} \simeq \frac{1}{4Af} \ln \left( 1 - 4Af\right).
\label{eq:crowding}
\end{equation}

\noindent Here $f$ denotes the {\it total} observed surface density of
images, $A$ is the average image area, and $f'$ is an estimate of the
true, {\it total} surface density. The average image area $A$ depends
on the threshold isophote above the local sky used  for
image detection. An image is defined as a region of simply connected
pixels exceeding this threshold, and the image size is consequently
affected by such factors as seeing and saturation effects. The
value of $A$ is typically a few times larger than the FWHM seeing disk
and is best determined by judicious experimentation (see below). If we
assume that, to first order, crowding effects are independent of
luminosity and color, then we may multiply our observed,
color-magnitude-biased surface densities by $\alpha_{c}$ to obtain an
estimate of the true stellar surface densities.

	How valid is this assumption? 	Equation \ref{eq:crowding} is
derived purely on the basis of identical, overlapping images and makes
no allowance for possible luminosity effects. It is reasonable to
suppose that in areas of high surface density a significant fraction
of relatively faint stars could be lost in the photon noise generated
by nearby bright stars.  Alternatively, faint star counts could be
enhanced in these areas owing to the underlying density of unresolved
objects and its effect on the threshold used for detection.
Consequently, it is of interest to examine our star counts as a
function of magnitude to determine the extent to which such effects
might be present.

	In Figures \ref{fig:bfratior} and \ref{fig:bfratiod} we show
the variation in the ratio of bright stars to faint stars with radius
and total surface density for a representative selection of clusters.
The clusters shown have all been observed under similar seeing
conditions (FWHM $\approx$2.5\arcsec). For each cluster we determine the
median magnitude for all images classified as stars and falling within
the appropriate color-magnitude envelope, then count the number of
stellar images above and below this magnitude.  It is apparent from
Figures \ref{fig:bfratior} and \ref{fig:bfratiod} that the scatter is
quite large and, more importantly, that there is considerable
variation in the run of the relative numbers of bright and faint stars
from cluster to cluster.  While the relative fractions of bright stars
are sometimes seen to rise near the cluster centers, the onset of this
rise does not appear to depend strongly on the total surface density.
This is no doubt due in part to the relative densities of field and
cluster stars, each having different intrinsic luminosity functions.
Interestingly, some clusters show almost no luminosity effects at all.

It should, in principle, be possible to account for the observed
luminosity functions of field and cluster stars and so attempt to
model consistent luminosity corrections. However, we note that the
magnitudes of the variations are relatively small. If we were disposed
towards correcting the surface densities by forcing the ratio of
bright stars to faint stars to be constant throughout the cluster,
then the most offending deviations in Figure \ref{fig:bfratiod} would
contribute a relatively modest 30\% to the counts.  This is smaller
than the computed crowding corrections at the same surface density. In
addition, we see in Figure \ref{fig:bfratior} that the ratios of
bright to faint stars are consistent with unity well within the
published tidal radii of these clusters. We consequently ignore
luminosity effects and consider the application of Equation
\ref{eq:crowding} directly.

	As previously noted, image classification becomes
problematic in regions of high surface density. In Figure
\ref{fig:contributions} we show the surface density profiles for
images classified as stars, merged images, and non-stellar sources for
NGC 7089. Also shown are the densities of merged images one would
expect from a straightforward application of Equation
\ref{eq:crowding}. The unreliability of the classification algorithm
is demonstrated by the rapid growth near the cluster center in the
number of images classified as non-stellar.  Assuming that the
underlying surface density of galaxies should be approximately uniform
across the plate, then the apparent overdensity of non-stellar objects
near the cluster center may reasonably be attributed to overlapping
stellar images. If the intensities of two overlapping stellar images
are moderately different, the fainter image may lose its identity and
the net, elongated profile will be classified as non-stellar.  The
$\otimes$s in Figure \ref{fig:contributions} result from combining the
densities of objects classified as merged with the densities of
non-stellar images in excess of the mean level computed well outside
the cluster ($r > 20'$). It is evident that, beyond the region where
multiple mergers become important, the number of such ``deemed''
merged images agrees quite well with the number predicted by Equation
\ref{eq:crowding}.

	Equation \ref{eq:crowding} is obviously quite sensitive to the
image size parameter $A$. Unfortunately, we have no reliable,
internally consistent method for determining its optimal value.  Plots
such as Figure \ref{fig:contributions} provide a rough guide, but
owing to complex multiple-merger and saturation effects in the inner
regions where we are most sensitive to $A$, their utility is
limited. Hence we are forced to rely on the star counts of previous
investigators. Specifically, we alter $A$ until our corrected surface
density profiles match as closely as possible the profiles of King
\etal's (1968) and Peterson's (1976) data in the inner regions.  The
correction is largest and most sensitive to $A$ at high surface
densities where King {\it et.al}'s counts are least susceptible to
background uncertainties, though we are then subject to the
uncertainties in their own crowding corrections. We also note that
equation \ref{eq:crowding} becomes undefined for $fA \geq 4$ and is
unlikely to be accurate near this regime. Nonetheless, matching star
count profiles by eye result in reasonable values for $A$ (Irwin \&
Trimble 1984), and the corrected surface densities shown in Figure
\ref{fig:sd} generally agree quite well with the counts of King \etal
over practically the entire range in common.

\section{Analysis. \label{sec:tidalradii} }

\subsection{Surface Density Profiles. \label{sec:sds} }

	Examination of Figure \ref{fig:sd} reveals that, while King
models generally provide an excellent match to the observations over a
large range in surface density, this agreement evidently breaks down
at some radius which varies from one cluster to the next. Most
clusters show an apparent excess of stars at $r \ge r_t$.  To better
illustrate the departures of the data from model predictions, we show
in Figure \ref{fig:o-c} the observed-minus-computed residuals (in the
log domain) using King models with core and tidal radii from Table 1.
The residuals are plotted against $r^{1/4}$ to better reveal the
observed structure at large radii.  We are more concerned here with
gradients in the residuals of individual data sets than with their
mean values (which rely on imperfect normalizations of each data set).
It is apparent that in {\it all} cases, both the star counts of King
\etal (1968) and those of the present study depart from the King
models at radii considerably less than $r_{t}$.  Moreover, the
residuals often show systematic trends (\ie non-zero gradients) even
before we reach the positive break-aways which we henceforth ascribe
to ``extra-tidal'' stars. NGC 5824 is a particularly extreme example
which clearly cannot be represented with a King model at any radius.
Note that luminosity effects and incompleteness due to image crowding
(Sec. \ref{sec:crowding}) will have {\it reduced} the measured surface
densities of cluster stars with respect to the surface density of
stars in the field; applying completeness corrections would serve only
to increase the apparent departures from King-like behavior.

	The slopes in the residuals corresponding to the outer
aperture photometry data are no doubt at least partially due to
uncertainties in the sky subtraction.  Note also that since the total
light will be dominated by the contributions of stars on the giant
branch but the star counts will be heavily weighted towards stars of
lower mass, the form of the residuals could, in some cases, be
attributable to mass segregation.  While we could investigate this
possibility using appropriate multi-mass King models, we are at
present less concerned with the internal structure of globular
clusters than we are with their limiting radii. If globular clusters
are limited in spatial extent by the action of Galactic tidal forces,
then we expect the removal of stars (and hence the limiting radii) to
be independent of mass.

	The outer surface density profiles (Figure \ref{fig:sd}) of
NGC 288, 1904, 5824, 7078 and NGC 7089 show the most significant
departures from King-like behavior. Deep plates, high cluster-star
surface densities, and low field-star surface densities make these
among our best observed clusters.  These departures resemble
power-laws with slopes which vary from -1.6 (NGC 288) to -5 (NGC
7078). The surface density profile of NGC 5824 is particularly
striking in that it appears to follow a power law over almost the
entire extent of the data, with some flattening at both small and
large radii.  With the exception of NGC 2808 (whose outskirts may be
heavily obscured - \eg Faulkner \etal 1992), all of our sample
clusters show evidence of at least some extra-tidal material.

	Could these extra-tidal extensions be spurious products of
inaccurate foreground subtraction? To estimate the effects of our
assumed form for the foreground distribution, we fit higher-order
surfaces to fields surrounding NGC 288 and NGC 7089.  The five-point
stars in Figures \ref{fig:sdhon288} and \ref{fig:sdhon7089} show the
results of fitting a third-order bivariate polynomial to the
foreground distributions in these fields (see also Figure
\ref{fig:n7089sky}).  While the amplitudes of the extensions are
somewhat reduced by virtue of having been modeled out, significant
differences between the counts and the King models remain.  Evidently,
complex and highly contrived models would be required to remove the
extensions completely. To test the possibility that non-uniformities
in the plates have led to local enhancements in the number of detected
images, we have reanalysed the data for these two clusters using only
the brightest 50\% of the stars within each color-magnitude envelope.
We find no significant differences between the resulting profiles and
those shown in Figure
\ref{fig:sd}.

We conclude that the extended profiles are not artifacts of our
counting procedures and are consequently real. To what then do we
attribute these extra-tidal extensions? Recent numerical simulations
by Oh and Lin (1992) and Grillmair \etal (1995) have shown that the
removal of loosely-bound cluster stars by the tidal field of the
Galaxy is a relatively inefficient process, and that clusters can be
surrounded by halos of unbound stars for many Galactic orbits.
Moreover, once removed from the cluster, unbound stars form tidal
tails extending over enormous distances ahead and behind the cluster
in its orbit. The volume densities within the tidal tails are subject
to the open-orbit analog to Kepler's 3rd Law; near apogalacticon, the
differential slowing of stars in the tidal tails cause the stars in
the tails to converge on the cluster. The projected surface densities
of tail-stars may increase by an order of magnitude over their values
at perigalacticon, depending on the shape of the cluster's orbit and
the orientation of the cluster along the line of sight.

 If we associate the extra-tidal material with extended halos of stars
in the process of leaving the cluster, then the assumed, King-like
surface density distribution of bound stars will be affected well
within the tidal radius by the column density of the extra-tidal stars
along the line of sight. The form of the surface density profile of
the extra-tidal material depends on the viewing angle, and it would
clearly be hazardous to attempt to model the volume density of
extra-tidal material based solely on the observed surface density
profile. If cluster distribution functions are truly King-like, then
an estimate of the tidal cutoff in the binding energy (which is
probably the determining factor in establishing which stars eventually
become unbound - Keenan (1981); Grillmair \etal 1995) requires that we
avoid fitting models to regions of the surface density profile
significantly contaminated by extra-tidal material.

	We determine the King-model tidal radius by estimating the
radius at which the surface density of extra-tidal material does not
contribute significantly to the total. This is clearly a rather
subjective endeavor, but the results are not very sensitive to our
chosen cutoff due to the distribution of weights, which strongly
favors data at small $r$. We estimate core radii by eye using the
aperture photometry and determine $r_t$ and the scaling constant from
our star counts using using a downhill simplex,
$\chi^{2}$-minimization routine. Our fitted values for $r_c$ and $r_t$
are listed in Table 16. The 90\% confidence intervals are
estimated by fitting several hundred Poisson realizations of the star
count data for each cluster.

	Having avoided our newly-discovered, extra-tidal extensions,
it is not surprising that our fitted tidal radii are generally very
similar to those of previous investigators. In several instances our
determination of $r_t$ is somewhat smaller than the corresponding
value in Table 1, a circumstance which we attribute to the inclusion
of unrecognized, extra-tidal material in previous analyses.  In view
of the power-law form evident in the extra-tidal surface density
profiles of some clusters (particularly NGC 288, NGC 1904, and NGC
7089), it is difficult to resist the temptation to fit the extensions
and the main body of data simultaneously. We have indeed done so, and
in some instances found the fitted, King tidal radii to be
considerably smaller than the published values, depending on the
relative weight of the data at large radii. However, the number of
parameters required to model the surface density distribution cannot
be justified given our limited knowledge of the distribution of
extra-tidal material in the vicinity of the clusters and the
orientation of the clusters with respect to our line of sight.

\subsection{Two-Dimensional Structure. \label{sec:2dst} }

	In panels $a$ of Figure \ref{fig:map} we show contour maps of
the smoothed, foreground-subtracted surface densities of images
classified as stars in each of our clusters. The surface densities,
initially computed over bins of 1.56$\times$1.56 arcmin, are
Fourier-smoothed with a Gaussian kernel with $\sigma$ = 16 arcmin.
The surface densities at $r < r_t$ are artificially depressed using a
cosine function prior to Fourier transforming to avoid overwhelming
the low-density structures beyond $r_t$. It is for this reason that
the innermost contours often are not centered precisely on the
cluster.  The heavy, circular contours arise from smoothing (in a
manner identical to that used for the real data) projected King models
having published core and tidal radii. The model surface densities
have been scaled as in Figure \ref{fig:sd}, and the contours shown
are numerically identical to the lowest contours of the real data
(shown as dashed lines). Also indicated in Figure \ref{fig:map} are
departures from zero of the foreground-subtracted surface densities
(in units of the standard deviation) in bins 25 arcminutes square. The
standard deviations have been computed from the actual scatter of the
mean surface densities of the 48 bins furthest from the cluster
center. The arrows at the positions of the cluster centers indicate
the direction to the Galactic center. Where more than one arrow is
shown, the longer of the two indicates the direction of the measured,
absolute proper motion of the cluster (Cudworth \& Hanson 1993).

The smoothed, contour maps show a combination of Poisson fluctuations
in the surface density of field stars, unrelated star clusters, and
what appear to be extensions of the clusters themselves.  How
significant are these extensions?  Could randomly distributed
overdensities of field stars masquerade as cluster extensions at this
level of smoothing? We have carried out simulations to determine the
relative power contained in the low order variations of the surface
densities and how this power compares with that expected from random
distributions of field stars.  Each simulation begins with a
128$\times$128 array containing the surface densities of all stars
meeting our color-magnitude criteria.  An appropriate bivariate
polynomial (\ie of the same order as that used in Section
\ref{sec:starcounts}) is fitted to the surface densities beyond the
tidal radius and subtracted.  The surface densities within the tidal
radius are left unaltered throughout the simulations since we wish to
examine only the power on large scales.  The fitted distribution of
foreground stars is binned into a 4$\times$4 array, from which are
computed the baseline number of stars in each of the 16 elements.  In
the course of each simulation, Poisson deviates are generated from
these baseline numbers and the appropriate number of stars are
randomly distributed across each element. The surface densities of
these stars are then modeled and subtracted exactly as above. In this
way we preserve any large-scale density gradients in the real data and
so simulate the additional effects of fitting and subtracting the
foreground distribution. Consider the case wherein strongly non-linear
gradients are present in the distribution of foreground stars (near
the Galactic plane or the SMC, for example).  Subtracting a low-order
function from the data would leave higher-order residuals which could
conceivably mask underlying extra-tidal cluster structure or, worse,
masquerade as cluster structure themselves.

	Each simulated, residual surface density map is then
Fast-Fourier-Transformed and the power spectral density recorded. A
total of 1000 simulations were carried out for each field. In panels
$b$ and $c$ of Figure \ref{fig:map} we compare the low-order power of
the real data with that of the simulations. The wavenumbers $k_{x}$
and $k_{y}$ correspond to the East-West and North-South directions,
respectively, and $k_{x},k_{y}$ = 0,0 represents the DC level.  In
panels $b$, the distributions of simulated power amplitudes (points)
are compared with those computed from the real data (filled circles)
and with those computed for isolated (\ie no surrounding field stars)
clusters (open circles). The mean values of the simulated
distributions are shown as open squares. In panels $c$ are shown the
fraction of simulated points with power amplitudes less than those
computed from the data. A value of 1.000 indicates that the power
amplitude computed from the data exceeds the highest value attained in
all 1000 realizations.

	It is apparent from these figures that the clusters with
noticeable extensions in panels $a$ often have low-order power spectra
significantly in excess of that resulting from a purely random
distribution of field stars. We can determine whether this power
corresponds to the large-scale features evident in panels $a$ by
reconstructing the surface density maps using only frequencies
containing excess power. In panels $d$ of Figure \ref{fig:map} are
shown surface density maps generated from the lowest 3 wavenumbers
with power amplitudes occurring in less than 0.1\% of the simulations.
(Obviously, if no excess power is evident, then an inverse Fourier
transform will result in an array of zeros - no contour maps are shown
in such cases).  The contour thresholds and increments are identical
to those used in panels $a$. For the case of NGC 288, we have removed
the cluster of images in the northwestern corner of the field before
carrying out these simulations. These reconstructions must be
interpreted with caution, but the similarities between panels $a$ and
$d$ indicate that, in several cases, the general form and scale the
extensions are significant at a high confidence level.

	While these tests are suggestive, they are by no means
conclusive. We have merely tested the hypothesis that Poisson
statistics alone could account for apparent large-scale structure.
Other factors which could contribute to excess power on large scales
include: ($i$) other clusters of stars in the field, ($ii$) uncertain
image classifications at faint magnitudes which could lead to the
inclusion of potentially significant numbers of galaxies in the
sample, and ($iii$) intervening clouds of obscuring material.

	In the case of NGC 6864, the relatively high counts extending
westwards from the cluster are  due to an unrelated
concentration of $\sim$ 200 disk stars. The concentration is
sufficient that we have been able to construct a rough
color-magnitude density map (Figure \ref{fig:n6864disksm}) as
described in Section \ref{sec:starcounts}. Comparing this map with the
color-magnitude morphology in the central regions of NGC 6864 (Figure
\ref{fig:cofm}) it is apparent that there are comparatively few
stars in the region corresponding to the turnoff of NGC 6864, and that
horizontal branch stars are not represented at all.  Hence we conclude
that the extension is not physically associated with NGC 6864. It is
unclear how far this ``cluster'' extends and, while we have
specifically avoided the western side of the cluster in our star
counts, we view the surface density profile in Figure \ref{fig:sd}
with some suspicion.

 For NGC 288, the high concentration of images in the northwestern
corner of the field is due to NGC 253, a giant, early-type spiral in the
Sculptor group. In most other cases, while localized concentrations of
stars may be apparent, the numbers of stars involved are insufficient
to allow us to say anything meaningful concerning their
color-magnitude morphology.  The contours extending southwards from
NGC 362 are suspect owing to the high-order surface fit required to
model the distribution of SMC stars.

	Given the uncertainties in star/galaxy separation discussed in
Section \ref{sec:foreground}, it is conceivable that the contours in
panels $a$ are influenced to some degree by background clusters of
galaxies. On the angular scales over which we are smoothing the
distributions of galaxies are intrinsically far more clumpy than the
distributions of foreground stars (MacGillivray \& Stobie 1984). To
estimate the extent to which stars have been misclassified by the APM
software, we have examined deep CCD images of the regions
$\approx$6\arcmin~ from the cores of NGC 7089 and NGC 288. These clusters
both show significant departures from King-like surface density
profiles, and the seeing conditions under which the plates were taken
were average, and slightly worse than average, respectively.  Images
fainter than APM $B_J = 20$ were inspected visually and were
classified as galaxies if they appeared to be significantly out of
round or underluminous in their cores. Of 42 images near NGC 288
classified by the APM as stellar, we found 6 which satisfied our
galaxy criteria.  1 out of 7 images classified as non-stellar appear
stellar in the CCD image. In NGC 7089, 11 out of 53
images classified by the APM as stellar have profiles which appear
galactic in character. Conversely, 5 out of 12 images classified as
non-stellar are quite stellar in appearance in our CCD images. The
conclusions we draw are that ($i$) some image confusion is indeed
present at faint magnitude levels, and ($ii$) stars and galaxies are
misclassified with approximately equal frequencies, with a possible
bias favoring galaxies.

In Figures \ref{fig:sdhon288} and \ref{fig:sdhon7089} are shown the
radial profiles of images classified as non-stellar for NGC 288 and
NGC 7089. The power-law profiles beyond $r_t$ apparent in the
(predominantly) stellar samples are not in evidence among the galaxy
counts. Indeed, the non-stellar image densities appear to some extent
to be anti-correlated with the total image counts, particularly where
the uncertainties in the star counts are relatively small.

	The clustering properties of galaxies are reasonably well
known (Peebles \& Hauser 1974; Heydon-Dumbleton \etal 1988). If the
contours in Figure \ref{fig:map} are significantly affected by
misclassified galaxies, then we should see the corresponding signature
in the two-point correlation function, $\omega(\theta)$. In Figure
\ref{fig:omega} we show the run of the ensemble estimator (Hewett
1982)

\begin{equation}
q\hat{\omega}(\theta) = \frac{\langle N_{i} N_{j} \rangle}
{\langle N_{i} \rangle \langle N_{j} \rangle} - 1
\label{eq:omega}
\end{equation}

\noindent for all images ({\it i.e.} of all colors and magnitudes)
surrounding NGC 7089 and classified as stellar and non-stellar,
respectively. The surface densities $N_{i}, N_{j}$ are computed within
cells of dimensions 3.1\arcmin$\times3.1$\arcmin~ and
foreground-subtracted with a 1$\times$1 bivariate polynomial (see
Figure \ref{fig:n7089sky}) to remove the field-star gradient. In
computing $\hat{\omega}$ we avoid the rectangular region extending
from 40\arcmin~ south to 55\arcmin~ north of the cluster wherein the
majority of postulated extra-tidal stars reside. The galaxies show the
expected power-law form, with an apparent break occurring near
0.5$^{\circ}$.  This is somewhat smaller than that found in galaxy
surveys and is most likely due to the depth and incompleteness of our
counts. The run of $\hat{\omega}$ for the stellar images is
gratifyingly close to zero on all scales, which is what one would
expect for a purely Poissonian distribution. This is also an
indication that our counts do not suffer unduly from plate or machine
measuring effects.

In panels $e$ of Figure \ref{fig:map} we show the smoothed
distributions of images classified as non-stellar, after removal of a
surface fit of the same order as that used for the stars. The contours
levels are numerically identical to the contour levels used in panels
$a$. The $1^{\circ}$ structure expected from galaxy counts is
apparent, but there is little detailed correspondence between this and
the structure apparent in panels $a$.  Indicated on these maps are the
linear correlation coefficients, $\xi$, determined by comparing the
sky-subtracted surface densities of stellar and non-stellar images in bins
25\arcmin~square. We have included in these calculations only the area
of the scan beyond the 30\arcmin~ so as not to be influenced by
variable crowding effects.  NGC 362, NGC 2808, and NGC 3201 show the
highest correlations. These are due to image classification
uncertainties induced by the high surface densities and to
differential obscuration by disk material.  Aside from these clusters,
NGC 7089 shows the highest correlation coefficient (0.17). In Figure
\ref{fig:sgcorr} we compare directly the sky-subtracted surface
densities of stellar and non-stellar images in the field surrounding
NGC 7089.  The strength of the correlation apparently relies most
heavily on a few outlying points and appears to be insignificant
otherwise.  The probability that 60 measurements of two uncorrelated
variables would give a correlation coefficient as large as 0.17 is
$\approx$21\%. We conclude from this and the above tests, apart from the
tree clusters mentioned above, that the contours in panels $a$ are not
significantly affected by misclassification of images.

	The question of obscuration is a more difficult one to
address. In the absence of image misclassification, we could in
principle determine the extent of the obscuration from the variation
in galaxy counts across the field. However, given the effects of image
crowding on misclassification of images and the likelihood that the
obscuring material is not confined to the region between us and the
cluster in question, any such attempt would be fraught with
uncertainties. We simply note that the surface density contours
surrounding NGC 2808 and NGC 3201 may be significantly affected by
obscuration, as indicated by the relatively high correlations between
the densities of stellar and non-stellar images. The
contours in Figure \ref{fig:map} for these clusters should be
interpreted with this in mind.

\section{Discussion \label{sec:discussion}}

	As noted above, the extra-tidal extensions in Figure
\ref{fig:sd} bear a striking resemblance to the halos and associated
tidal tails discussed by Oh and Lin (1992) and Grillmair \etal (1995).
The extensions become apparent at a surface density between 4 and 6
orders of magnitude below the central surface density, consistent with
the N-body simulations.  In Figure \ref{fig:n7089nbod} we compare the
surface density profiles of NGC 7089 and Grillmair \etal's model 64eoa
near its 14th perigalacticon. This model cluster was evolved along an
eccentric orbit with perigalactic distance of 2.5 kpc and apogalactic
distance of 10 kpc in a logarithmic potential with circular velocity
of 220 km s$^{-1}$. The model units have been scaled so that $r_h = 10
pc$ and the cluster appears as it would from a distance of 12 kpc
(Peterson 1993). The viewing angle is chosen to be along the line
connecting the cluster and the galactic center. Grillmair \etal (1995)
used a Jaffe model as their starting point in the simulations, giving
rise to the evidently non-King-like behavior of the surface density
profile in Figure \ref{fig:n7089nbod}. Nonetheless, the agreement
between the form of the model profile and the data beyond $r_t$ is
intriguing.

	Grillmair \etal (1995) find that both the amplitude and the
slope of the surface density profile beyond the tidal radius are a
function of the shape of the cluster orbit, the orbital phase of the
cluster, and the viewing angle. There are too many parameters
available to put meaningful constraints on the possible orientation
and orbital phase of NGC 7089.  We simply note that the length scales
and density contrasts apparent in the observations are consistent with
results of numerical simulations.  Extra-tidal stars have been found
in dwarf spheroidal satellites of the Milky Way Galaxy (Irwin \&
Hatzidimitriou 1993). That tidal tails in the near vicinity of our
sample clusters should still be detectable at the present epoch is
presumably a consequence of continual tidal shocking, disk shocking,
tidal torquing, and ongoing 2-body encounters in the cores.  The
diversity in the radial gradients of the extra-tidal material among
the clusters we have examined is consistent with the orbital shape and
phase dependence revealed in the numerical simulations.

	In Figure \ref{fig:contnbod} we show contour plots of the
two-dimensional surface density distribution of Grillmair \etal's
(1995) model 64eoa at its sixth orbital apocenter as it would appear
from a distance of 12 kpc from three different perspectives. The
surface densities have been smoothed with the same kernel and masking
radius used for NGC 7089 in Figure \ref{fig:map}. The forms of the
surface density distributions are consistent to a greater or lesser
extent with many of the contour maps in Figure \ref{fig:map}.  In the
case of NGC 7089, Figure \ref{fig:map} indicates that the absolute
proper motion is directed almost perpendicularly to the apparent major
axis of the extra-tidal material. Though this might seem surprising,
inspection of Figure \ref{fig:contnbod} suggests that we may be seeing
NGC 7089 near it's apogalacticon and that the angle between the plane
of the cluster's orbit and our line of sight is relatively large. The
latter would be consistent with the Galactic coordinates and derived
space motion of this cluster (Cudworth \& Hanson 1993). The magnitude
of the derived space motion ($\approx 300$ km s$^{-1}$ in the Galactic
rest frame) is rather large, even if the cluster is assumed to be on a
circular orbit, though the uncertainties are substantial. The space
motion derived for NGC 7078 indicates that it's orbital plane is more
highly inclined to our line of sight, consistent with the more
``in-line'' appearance of it's extra-tidal material. For the remaining
clusters, the lack of velocity information and the poor statistics of
the observed surface density distributions conspire to make
distinctions concerning probable viewing angles largely meaningless.

We conclude that the stars we find beyond the best-fit values of $r_t$
are probably unbound as a result of previous and ongoing stripping
episodes.  Given that our highest quality data show quite pronounced
extra-tidal extensions, and that the majority of clusters in our
sample appear to have at least some extra-tidal material, we speculate
that globular clusters in general have no observable limiting radii.
Without {\it a priori} knowledge of the spatial distribution of
unbound stars in the vicinity of a cluster, the limits imposed by the
Galactic tidal field on the spatial distribution of {\it bound} stars
can only be estimated by model-dependent means. If the underlying
distribution of bound stars is King-like, then the tidal cutoff in the
binding energy can best be determined by fitting models only to data
not significantly affected by the presence of unbound stars. In
effect, the presence of unbound stars places an upper limit on the
accuracy to which we can resolve the tidal radii of globular clusters.
Despite the quality of our data, our ability to determine the
distribution of globular cluster orbit shapes (or conversely, the
Galactic mass distribution) is little better than that of previous
investigators (Peterson 1974; Innanen, Harris, \& Webbink 1983).

Some of the smoothed surface density contour maps in Figure
\ref{fig:map} are suggestive of tidal tails extending to considerable
distances from the clusters. To verify that these tails are physically
associated with the clusters will require spectroscopy of a reasonable
sample of extra-tidal stars with colors and magnitudes consistent with
those of the clusters. If their velocities and metallicities are
similar to those of bound cluster stars, we may be reasonably
confident that the tails are not due to other stellar associations in
the field.  Sufficient velocity resolution would be required to
distinguish between cluster and halo stars. On the other hand, the
radial velocities of the extra-tidal stars at some distance from the
cluster may be expected to differ appreciably from that of the cluster
itself, depending on the shape of the cluster's orbit, the orbital
phase of the cluster, and the orientation with respect to our line of
sight. As an example, we show in Figure \ref{fig:radialv} the
line-of-sight velocities for particles in the model used to construct
Figure \ref{fig:contnbod}. The velocities shown have been projected
along the line connecting the cluster to the galactic center, after
subtracting the motion of the cluster itself. While the velocity
variations across the field may or may not be large depending on the
variables listed above, the velocity {\it dispersions} of the stars at
any point within the tidal tails will be comparatively small.  Since
most of our observed extra-tidal stars have $B_J > 20$ and the numbers
of tail stars are not large to begin with, a program to
spectroscopically confirm the association of these stars with their
parent clusters would entail considerable effort, presumably requiring
multi-object spectrographs on 4-meter-class telescopes.

	Given the great ages of globular clusters, we expect that the
tidal tails we have found should extend well beyond the 4$^{\circ}$
fields we have examined. For NGC 7089 we find that the one-dimensional
surface density profile beyond $r_t$ goes as $f \propto r^{-2}$,
implying that the two-dimensional surface density is proportional to
$r^{-1}$. It may be possible to detect unbound cluster stars at even
larger distances from the cluster than we have examined here. The
appearance of the tidal tails on large scales is determined primarily
by the cluster's orbit shape and less by the orbital phase of the
cluster or the details of the stripping process. Defining the
large-scale distribution of extra-tidal stars on the sky would, in
principle, allow us to actually trace the cluster's orbit. Estimates
of the tidal radii of globular clusters may not be as useful for
studying Galactic dynamics as we had hoped, but it may yet be possible
to determine orbit shapes of clusters from the distributions and
velocities of their leavings.

\acknowledgments

We thank the staff of the United Kingdom Schmidt Telescope Unit for
their untiring efforts to acquire the plate material used in this
investigation. CJG acknowledges the support of an Australian National
University PhD scholarship.

\clearpage

\begin{figure}[h]
\caption{APM color-magnitude diagrams of all images labeled as stars
for each of our clusters. The left-hand panels show the
color-magnitude distribution of stars lying typically between 0.1 and
$0.5r_t$, and the right-hand panels show the distribution of field
stars. The heavy lines indicate the regions chosen for counting
purposes to give the highest signal-to-noise ratios in the outskirts
of each cluster.}
\label{fig:cofm}
\end{figure}

\begin{figure}[h]
\caption{Contour plot of the smoothed density distribution of cluster
stars in NGC 3201. The thin lines map the density distribution of
cluster stars as per Equation 1. The heavy line is a contour
of constant signal-to-noise ratio (computed using Equation
2 and subsequently smoothed) arbitrarily chosen to match
the density distribution on the red side of the cluster sequence.}
\label{fig:n3201cmmap}
\end{figure}

\begin{figure}[h]
\caption{Cumulative star-count signal-to-noise ratio $S$ as a function
of enclosed color-magnitude area, $a_{k}$, for NGC 362.}
\label{fig:n362snvsarea}
\end{figure}

\begin{figure}[h]
\caption{Comparison of 1st and 3rd order bivariate polynomial fits to
the foreground distribution surrounding NGC 7089. The regions excluded
from the surface fit are indicated by heavy lines.}
\label{fig:n7089sky}
\end{figure}

\begin{figure}[h]
\caption{The software r\'{e}seau used for the star counts in Tables 3
through 14. Sector numbers are indicated.}
\label{fig:sectors}
\end{figure}

\begin{figure}[h]
\caption{Surface density profiles of our sample clusters. Hollow
symbols refer to aperture photometry, filled symbols represent
crowding-corrected star count data, and the $\times$s indicate our
star counts uncorrected for crowding. References to the original
sources of data are given in Table 15. The solid lines are King
models, normalized to our star-count data, with core and tidal radii
taken from Table 1. The run of field star surface densities (corrected
for crowding effects) are shown by dotted lines.}
\label{fig:sd}
\end{figure}

\begin{figure}[h]
\caption{Ratio of bright stars to faint stars as a function of radius
in a representative selection of sample clusters.}
\label{fig:bfratior}
\end{figure}

\begin{figure}[h]
\caption{Ratio of bright stars to faint stars as a function of total
surface density in a representative selection of sample clusters.}
\label{fig:bfratiod}
\end{figure}

\begin{figure}[h]
\caption{Contributions to the total surface density by
different classes of images.  $\otimes$s represent
images ``deemed'' merged and include the excess surface density of
non-stellar images at small radii. The curve shows the expected
density of merged objects based on Equation 7 and an
image area of 34 arcsec$^{2}$.}
\label{fig:contributions}
\end{figure}

\begin{figure}[h]
\caption{Residuals after subtracting (in the log domain)
normalized King models from the surface density data. Arrows indicate
core and tidal radii from Table 1. References to
sources of data are given in Table 15.}
\label{fig:o-c}
\end{figure}

\begin{figure}[h]
\caption{Surface density profiles for NGC 288 computed
using both $1 \times 1$ and $3 \times 3$ fits to the surface densities
of foreground stars. Also shown are the surface densities computed for
objects classified as non-stellar, using a $1 \times 1$ fit to the
distribution of field objects. The solid curve is a King model with
core and tidal radii from Table 1.}
\label{fig:sdhon288}
\end{figure}

\begin{figure}[h]
\caption{Same as Figure 11, but for NGC 7089.}
\label{fig:sdhon7089}
\end{figure}

\begin{figure}[h]
\caption{Contour maps of the background-subtracted and
Fourier-smoothed surface density distributions. Panels $a$ show the
contours of images classified as stars after Fourier smoothing with a
Gaussian kernel of 16\arcmin.  The inner regions of the clusters have
been cosbelled out to reveal structure at large radii. Boxed numbers
indicate deviations from zero in units of the standard deviation, and
the heavy, circular contours arise from smoothing King-model surface
density distributions in an identical manner. The numerical values of
these heavy contours are identical to the lowest (dashed) contours of
the data. Arrows at the centers of the clusters indicate the direction
to the Galactic center.  Where more than one arrow is shown, the
longer of the two indicates the direction of the measured, absolute
proper motion of the cluster.  Panels $b$ and $c$ show the results of
1000 Poissonian realizations of each field, as described in the text.
For those cases in which excess power on large scales is evident,
panels $d$ show the smoothed contours of just the lowest three
wavenumbers with power amplitudes significant at $>$99\% confidence
level. Panels $e$ show similarly smoothed contours of the surface
density distribution of images classified as non-stellar. For each
cluster, the lowest (dashed) contours and the contour intervals in
panels $a, d$, and $e$ are identical.}
\label{fig:map}
\end{figure}

\begin{figure}[h]
\caption{Color-magnitude distribution of images extending westwards
from NGC 6864.}
\label{fig:n6864disksm}
\end{figure}

\begin{figure}[h]
\caption{Two-point correlation function for stellar (filled circles)
and non-stellar (open circles) images surrounding NGC 7089. Error bars
are shown for alternating points to avoid confusion.}
\label{fig:omega}
\end{figure}

\begin{figure}[h]
\caption{Comparison of the foreground-subtracted surface densities of
non-stellar images with those of stellar images for the field
surrounding NGC 7089.}
\label{fig:sgcorr}
\end{figure}

\clearpage

\begin{figure}[h]
\caption{Comparison between a subset of stars counts and aperture
photometry for NGC 7089 with the surface density profile of model
64eoa of Grillmair \etal (1995) after 14 orbits. The solid line
corresponds to a King model with core and tidal radii taken from Table
1, and the dashed line shows the surface density profile of our model
as it would appear from a distance of 12 kpc looking out from the
galactic center. Codes for data sources are given in Table 15.}
\label{fig:n7089nbod}
\end{figure}

\begin{figure}[h]
\caption{Smoothed, 2-dimensional surface density distribution of
particles in model 64eoa of Grillmair \etal (1995) after six
perigalactic passages, as viewed from a distance of 12 kpc along a
line connecting the cluster and the Galactic center (top), along a
tangent to the orbital path (middle), and along a line perpendicular
to the orbital plane (bottom).  The cluster's direction of motion is
indicated in each case.}
\label{fig:contnbod}
\end{figure}

\begin{figure}[h]
\caption{Particle velocities, projected along a line connecting the
cluster and the galactic center, for the model shown smoothed in
Figure 18. The model used is near it's orbital apogalacticon so that
the tidal tails extend nearly perpendicularly to the line of sight. The
projected distances from the cluster center have been computed
assuming a distance of 12 kpc between the cluster and the observer.}
\label{fig:radialv}
\end{figure}

\end{document}


\begin{planotable}{lrrrrrrr}
\tablecaption{Relevant Cluster Data.}
\tablewidth{15cm}
\tablehead{
\colhead{Cluster} &
\colhead{$\alpha$\tablenotemark{a}} &
\colhead{$\delta$\tablenotemark{a}} &
\colhead{$l^{II}$} &
\colhead{$b^{II}$} &
\colhead{$r_c$\tablenotemark{b}} &
\colhead{$r_t$\tablenotemark{b}} &
\colhead{$R_{GC}\tablenotemark{c}$}\nl
\colhead{} &
\colhead{(1950)} &
\colhead{(1950)} &
\colhead{($^{\circ}$)} &
\colhead{($^{\circ}$)} &
\colhead{(arcmin)} &
\colhead{(arcmin)} &
\colhead{(kpc)}}

\startdata

NGC 288  & 00 50 21 & -26 51 41 & 152.2 & -89.4 & 1.42 & 12.9 & 11.7\nl
NGC 362  & 01 01 33 & -71 06 59 & 301.5 & -46.2 & 0.17 &
14.9 & 9.3\nl
NGC 1904 & 05 22 07 & -24 34 08 & 227.2 & -29.4 & 0.16 &
8.4 &19.1\nl
NGC 2808 & 09 11 04 & -64 39 23 & 282.2 & -11.3 & 0.26 &
15.6 & 11.0\nl
NGC 3201 & 10 15 34 & -46 09 38 & 227.2 & 8.6 &
1.45 & 29.6 & 8.9\nl
NGC 4590 & 12 36 49 & -26 28 06 & 299.6 & 36.1 & 0.69 &
30.3 & 9.6\nl
NGC 5824 & 15 00 54 & -32 52 23 & 332.6 & 22.1 & 0.06 &
15.6 & 26.3\nl
NGC 6864 & 20 03 08 & -22 03 55 & 20.3 & -25.7 &
0.10 & 7.3 & 12.6\nl
NGC 6934 & 20 31 44 & 07 13 55 & 52.1 & -18.9 &
0.25 & 8.4 & 12.0\nl
NGC 6981 & 20 50 43 & -12 43 37 & 35.2 & -32.7 &
0.54 & 9.2 & 13.8\nl
NGC 7078 & 21 27 33 & 11 56 49 & 65.0 & -27.3 &
0.07 & 21.5 & 10.5\nl
NGC 7089 & 21 30 55 & -01 02 44 & 53.4 & -35.8 &
0.34 & 21.5 & 10.7\nl\nl
\vspace{-4.0ex}
\tablecomments{}
\tablenotetext{a}{Shawl \& White (1986).}
\tablenotetext{b}{Trager, Djorgovski, \& King (1993).}
\tablenotetext{c}{Djorgovski (1993).}
\label{tab:data}
\end{planotable}
\clearpage

\begin{small}
\begin{planotable}{lcccrrl}
\tablecaption{Observations.}
\tablewidth{17cm}
\tablehead{
\colhead{Cluster} &
\colhead{Plate ID\#} &
\colhead{Emulsion} &
\colhead{Filter} &
\colhead{Exposure Time} &
\colhead{Image Diameter} &
\colhead{} \nl
\colhead{} &
\colhead{} &
\colhead{} &
\colhead{} &
\colhead{minutes} &
\colhead{microns} &
\colhead{}}

\startdata

NGC 288  & 13390 & IIIaF & RG 630 & 105 & 50&\nl
	 & 13437 & IIIaF & RG 630 & 90 & 40&\nl
         & 13840 & IIIaJ & GG 395 & 65 & 45&\nl
         & 13850 & IIIaJ & GG 395 & 65 & 35&\nl
\nl
NGC 362  & 13276 & IIIaF & RG 630 & 45 & 25&\nl
	 & 13277 & IIIaJ & GG 395 & 45 & 45&\nl
	 & 14000 & IIIaF & RG 630 & 45 & 35&\nl
         & 13999 & IIIaJ & GG 395 & 30 & 35&\nl
\nl
NGC 1904 & 13421 & IIIaJ & GG 395 & 65 & 35&\nl
         & 13432 & IIIaJ & GG 395 & 70 & 30&\nl
         & 13453 & IIIaF & RG 630 & 90 & 35&\nl
         & 13455 & IIIaF & RG 630 & 85 & 50&\nl
\nl
NGC 2808 & 13482 & IIIaJ & GG 395 & 25 & 25&\nl
         & 13509 & IIIaF & RG 630 & 35 & 35&\nl
         & 13542 & IIIaJ & GG 395 & 25 & 50&\nl
         & 13543 & IIIaF & RG 630 & 35 & 35&\nl
\nl
NGC 3201 & 13549 & IIIaJ & GG 395 & 25 & 30&\nl
         & 13550 & IIIaF & RG 630 & 35 & 35&\nl
         & 13561 & IIIaJ & GG 395 & 25 & 30&\nl
         & 13562 & IIIaF & RG 630 & 35 & 35&\nl
\nl
NGC 4590 & 13606 & IIIaF & RG 630 & 90 & 40&\nl
         & 13737 & IIIaJ & GG 395 & 60 & 35&\nl
         & 13754 & IIIaF & RG 630 & 90 & 50&\nl
         & 13756 & IIIaJ & GG 395 & 60 & 35&\nl
\nl
NGC 5824 & 13615 & IIIaJ & GG 395 & 65 & 35&\nl
         & 13621 & IIIaJ & GG 395 & 60 & 40&\nl
         & 13666 & IIIaF & RG 630 & 90 & 30&\nl
         & 13673 & IIIaF & RG 630 & 90 & 35&\nl
\nl
NGC 6864 & 13231 & IIIaJ & GG 395 & 25 & 40&\nl
         & 13240 & IIIaF & RG 630 & 35 & 40&\nl
         & 13674 & IIIaJ & GG 395 & 25 & 35&\nl
         & 13654 & IIIaF & RG 630 & 45 & 45&\nl
\nl
NGC 6934 & 13725 & IIIaJ & GG 395 & 60 & 35&\nl
         & 13813 & IIIaF & RG 630 & 90 & 35&\nl
         & 13818 & IIIaJ & GG 395 & 70 & 35&\nl
\nl
NGC 6981 & 13248 & IIIaF & RG 630 & 120 & 35&\nl
         & 13274 & IIIaJ & GG 395 & 65 & 35&\nl
         & 13387 & IIIaF & RG 630 & 105 & 35&\nl
         & 13829 & IIIaJ & GG 395 & 65 & 40&\nl
\nl
NGC 7078 & 13339 & IIIaF & RG 630 & 100 & 60&\nl
         & 13353 & IIIaJ & GG 395 & 70 & 50&\nl
	 & 13389 & IIIaF & RG 630 & 105 & 50&\nl
         & 13726 & IIIaJ & GG 395 & 60 & 35&\nl
\nl
NGC 7089 & 13243 & IIIaJ & GG 395 & 60 & 35&\nl
         & 13348 & IIIaF & RG 630 & 100 & 35&\nl
         & 13732 & IIIaF & RG 630 & 90 & 35&\nl
         & 13855 & IIIaJ & GG 395 & 65 & 35&\nl

\label{tab:obs}
\end{planotable}
\end{small}
\begin{small}
\begin{planotable}{lcccrrl}
\tablecaption{Observations.}
\tablewidth{17cm}
\tablehead{
\colhead{Cluster} &
\colhead{Plate ID\#} &
\colhead{Emulsion} &
\colhead{Filter} &
\colhead{Exposure Time} &
\colhead{Image Diameter} &
\colhead{} \nl
\colhead{} &
\colhead{} &
\colhead{} &
\colhead{} &
\colhead{minutes} &
\colhead{microns} &
\colhead{}}

\startdata

NGC 288  & 13390 & IIIaF & RG 630 & 105 & 50&\nl
	 & 13437 & IIIaF & RG 630 & 90 & 40&\nl
         & 13840 & IIIaJ & GG 395 & 65 & 45&\nl
         & 13850 & IIIaJ & GG 395 & 65 & 35&\nl
\nl
NGC 362  & 13276 & IIIaF & RG 630 & 45 & 25&\nl
	 & 13277 & IIIaJ & GG 395 & 45 & 45&\nl
	 & 14000 & IIIaF & RG 630 & 45 & 35&\nl
         & 13999 & IIIaJ & GG 395 & 30 & 35&\nl
\nl
NGC 1904 & 13421 & IIIaJ & GG 395 & 65 & 35&\nl
         & 13432 & IIIaJ & GG 395 & 70 & 30&\nl
         & 13453 & IIIaF & RG 630 & 90 & 35&\nl
         & 13455 & IIIaF & RG 630 & 85 & 50&\nl
\nl
NGC 2808 & 13482 & IIIaJ & GG 395 & 25 & 25&\nl
         & 13509 & IIIaF & RG 630 & 35 & 35&\nl
         & 13542 & IIIaJ & GG 395 & 25 & 50&\nl
         & 13543 & IIIaF & RG 630 & 35 & 35&\nl
\nl
NGC 3201 & 13549 & IIIaJ & GG 395 & 25 & 30&\nl
         & 13550 & IIIaF & RG 630 & 35 & 35&\nl
         & 13561 & IIIaJ & GG 395 & 25 & 30&\nl
         & 13562 & IIIaF & RG 630 & 35 & 35&\nl
\nl
NGC 4590 & 13606 & IIIaF & RG 630 & 90 & 40&\nl
         & 13737 & IIIaJ & GG 395 & 60 & 35&\nl
         & 13754 & IIIaF & RG 630 & 90 & 50&\nl
         & 13756 & IIIaJ & GG 395 & 60 & 35&\nl
\nl
NGC 5824 & 13615 & IIIaJ & GG 395 & 65 & 35&\nl
         & 13621 & IIIaJ & GG 395 & 60 & 40&\nl
         & 13666 & IIIaF & RG 630 & 90 & 30&\nl
         & 13673 & IIIaF & RG 630 & 90 & 35&\nl
\nl
NGC 6864 & 13231 & IIIaJ & GG 395 & 25 & 40&\nl
         & 13240 & IIIaF & RG 630 & 35 & 40&\nl
         & 13674 & IIIaJ & GG 395 & 25 & 35&\nl
         & 13654 & IIIaF & RG 630 & 45 & 45&\nl
\nl
NGC 6934 & 13725 & IIIaJ & GG 395 & 60 & 35&\nl
         & 13813 & IIIaF & RG 630 & 90 & 35&\nl
         & 13818 & IIIaJ & GG 395 & 70 & 35&\nl
\nl
NGC 6981 & 13248 & IIIaF & RG 630 & 120 & 35&\nl
         & 13274 & IIIaJ & GG 395 & 65 & 35&\nl
         & 13387 & IIIaF & RG 630 & 105 & 35&\nl
         & 13829 & IIIaJ & GG 395 & 65 & 40&\nl
\nl
NGC 7078 & 13339 & IIIaF & RG 630 & 100 & 60&\nl
         & 13353 & IIIaJ & GG 395 & 70 & 50&\nl
	 & 13389 & IIIaF & RG 630 & 105 & 50&\nl
         & 13726 & IIIaJ & GG 395 & 60 & 35&\nl
\nl
NGC 7089 & 13243 & IIIaJ & GG 395 & 60 & 35&\nl
         & 13348 & IIIaF & RG 630 & 100 & 35&\nl
         & 13732 & IIIaF & RG 630 & 90 & 35&\nl
         & 13855 & IIIaJ & GG 395 & 65 & 35&\nl

\label{tab:obs}
\end{planotable}
\end{small}
\clearpage

\begin{planotable}{crrrrrr}
\tablewidth{265pt}
\tablecaption{NGC 288 Star Counts.}
\tablehead{
\colhead{$r$} &
\colhead{$r_{eff}$} &
\colhead{N} &
\colhead{f$_{bkd}$} &
\colhead{$\alpha_c$} &
\colhead{f} &
\colhead{Error}}

\startdata

 0 -   1&  0.71&    17&  0.152&\nodata&\nodata&\nodata\nl
 1 -   2&  1.58&    86&  0.152&\nodata&\nodata&\nodata\nl
 2 -   3&  2.55&   209&  0.152&\nodata&\nodata&\nodata\nl
 3 -   4&  3.54&   401&  0.152& 3.923&  71.381&   3.572\nl
 4 -   5&  4.53&   463&  0.152& 1.705&  27.774&   1.298\nl
 5 -   6&  5.52&   400&  0.152& 1.341&  15.373&   0.776\nl
 6 -   7&  6.52&   291&  0.152& 1.172&   8.196&   0.489\nl
 7 -   8&  7.52&   229&  0.152& 1.113&   5.256&   0.357\nl
 8 -   9&  8.51&   141&  0.152& 1.064&   2.656&   0.236\nl
 9 -  10&  9.51&    96&  0.152& 1.038&   1.517&   0.170\nl
10 -  12& 11.05&   100&  0.152& 1.023&   0.588&   0.074\nl
12 -  14& 13.04&    78&  0.152& 1.020&   0.335&   0.055\nl
14 -  16& 15.03&    44&  0.152& 1.014&   0.085&   0.036\nl
16 -  18& 17.03&    42&  0.152& 1.015&   0.047&   0.031\nl
18 -  20& 19.03&    49&  0.152& 1.014&   0.056&   0.030\nl
20 -  25& 22.64&   118&  0.152& 1.014&   0.017&   0.016\nl
25 -  30& 27.61&   134&  0.152& 1.015&   0.005&   0.014\nl
30 -  35& 32.60&   182&  0.152& 1.016&   0.029&   0.013\nl
35 -  40& 37.58&   186&  0.152& 1.017&   0.008&   0.012\nl
40 -  50& 45.28&   477&  0.152& 1.017&   0.019&   0.008\nl
50 -  60& 55.23&   531&  0.152& 1.017&   0.004&   0.007\nl
60 -  70& 65.19&   607&  0.152& 1.016&  -0.001&   0.006\nl
70 -  80& 75.17&   717&  0.152& 1.017&   0.003&   0.006\nl
80 -  90& 85.15&   814&  0.152& 1.016&   0.003&   0.005\nl
90 - 100& 95.13&   884&  0.152& 1.016&  -0.002&   0.005\nl
\nl
\multicolumn{4}{r}{Sectors used:} & \multicolumn{3}{l}{1-8}\nl
\multicolumn{4}{r}{Surface Fit Order:} &
\multicolumn{3}{l}{1$\times$1}\nl
\multicolumn{4}{r}{Image Area $A$:} & \multicolumn{3}{l}{27.2 arcsec$^{2}$}
\nl

\end{planotable}

\clearpage

\begin{planotable}{crrrrrr}
\tablewidth{265pt}
\tablecaption{NGC 362 Star Counts.}
\tablehead{
\colhead{$r$} &
\colhead{$r_{eff}$} &
\colhead{N} &
\colhead{f$_{bkd}$} &
\colhead{$\alpha_c$} &
\colhead{f} &
\colhead{Error}}

\startdata

 0 -   1&  0.71&     0&  0.848&\nodata&\nodata&\nodata\nl
 1 -   2&  1.58&    20&  0.841&\nodata&\nodata&\nodata\nl
 2 -   3&  2.55&   139&  0.829&\nodata&\nodata&\nodata\nl
 3 -   4&  3.54&   281&  0.816&\nodata&\nodata&\nodata\nl
 4 -   5&  4.53&   277&  0.805& 2.551&  39.183&   2.403\nl
 5 -   6&  5.52&   256&  0.795& 1.605&  18.226&   1.189\nl
 6 -   7&  6.52&   162&  0.784& 1.261&   7.218&   0.629\nl
 7 -   8&  7.52&   142&  0.773& 1.213&   5.073&   0.491\nl
 8 -   9&  8.51&    99&  0.764& 1.150&   2.646&   0.343\nl
 9 -  10&  9.51&    73&  0.754& 1.120&   1.438&   0.257\nl
10 -  12& 11.05&   105&  0.739& 1.102&   0.600&   0.131\nl
12 -  14& 13.04&   102&  0.721& 1.097&   0.375&   0.109\nl
14 -  16& 15.03&   112&  0.704& 1.087&   0.330&   0.098\nl
16 -  18& 17.03&   121&  0.687& 1.084&   0.295&   0.089\nl
18 -  20& 19.03&    86&  0.672& 1.072&  -0.054&   0.067\nl
20 -  25& 22.64&   270&  0.647& 1.076&   0.011&   0.040\nl
25 -  30& 27.61&   302&  0.615& 1.074&  -0.015&   0.035\nl
30 -  35& 32.60&   364&  0.588& 1.070&   0.023&   0.032\nl
35 -  40& 37.58&   377&  0.563& 1.067&  -0.017&   0.028\nl
40 -  50& 45.28&   924&  0.532& 1.063&   0.024&   0.018\nl
50 -  60& 55.23&  1008&  0.494& 1.061&   0.001&   0.016\nl
60 -  70& 65.19&  1069&  0.457& 1.055&  -0.015&   0.014\nl
70 -  80& 75.17&  1124&  0.415& 1.053&  -0.013&   0.012\nl
80 -  90& 85.15&  1176&  0.361& 1.049&   0.009&   0.011\nl
\nl
\multicolumn{4}{r}{Sectors used:} & \multicolumn{3}{l}{1-5}\nl
\multicolumn{4}{r}{Surface Fit Order:} &
\multicolumn{3}{l}{3$\times$2}\nl
\multicolumn{4}{r}{Image Area $A$:} & \multicolumn{3}{l}{27.2 arcsec$^{2}$}
\nl

\end{planotable}
\clearpage

\begin{planotable}{crrrrrr}
\tablewidth{265pt}
\tablecaption{NGC 1904 Star Counts.}
\tablehead{
\colhead{$r$} &
\colhead{$r_{eff}$} &
\colhead{N} &
\colhead{f$_{bkd}$} &
\colhead{$\alpha_c$} &
\colhead{f} &
\colhead{Error}}

\startdata

 0 -   1&  0.71&     0&  0.276&\nodata&\nodata&\nodata\nl
 1 -   2&  1.58&    91&  0.276&\nodata&\nodata&\nodata-\nl
 2 -   3&  2.55&   345&  0.276& 3.069&  67.128&   3.629\nl
 3 -   4&  3.54&   360&  0.276& 1.510&  24.442&   1.303\nl
 4 -   5&  4.53&   224&  0.276& 1.177&   9.048&   0.623\nl
 5 -   6&  5.52&   129&  0.276& 1.090&   3.794&   0.358\nl
 6 -   7&  6.52&    86&  0.276& 1.059&   1.954&   0.240\nl
 7 -   8&  7.52&    84&  0.276& 1.044&   1.586&   0.203\nl
 8 -   9&  8.51&    46&  0.276& 1.036&   0.616&   0.132\nl
 9 -  10&  9.51&    39&  0.276& 1.034&   0.400&   0.108\nl
10 -  12& 11.05&    67&  0.276& 1.031&   0.224&   0.061\nl
12 -  14& 13.04&    73&  0.276& 1.030&   0.185&   0.054\nl
14 -  16& 15.03&    66&  0.276& 1.026&   0.084&   0.044\nl
16 -  18& 17.03&    78&  0.276& 1.029&   0.100&   0.043\nl
18 -  20& 19.03&    72&  0.276& 1.030&   0.035&   0.037\nl
20 -  25& 22.64&   222&  0.276& 1.028&   0.047&   0.022\nl
25 -  30& 27.61&   223&  0.276& 1.027&  -0.010&   0.018\nl
30 -  35& 32.60&   270&  0.276& 1.027&  -0.004&   0.017\nl
35 -  40& 37.58&   288&  0.276& 1.025&  -0.025&   0.015\nl
40 -  50& 45.28&   736&  0.276& 1.027&  -0.008&   0.010\nl
50 -  60& 55.23&   897&  0.276& 1.026&  -0.009&   0.009\nl
60 -  70& 65.19&  1120&  0.276& 1.027&   0.006&   0.008\nl
70 -  80& 75.17&  1287&  0.276& 1.027&   0.005&   0.008\nl
80 -  90& 85.15&  1449&  0.276& 1.026&   0.003&   0.007\nl
90 - 100& 95.13&  1611&  0.276& 1.027&   0.002&   0.007\nl

\nl
\multicolumn{4}{r}{Sectors used:} & \multicolumn{3}{l}{1-8}\nl
\multicolumn{4}{r}{Surface Fit Order:} &
\multicolumn{3}{l}{1$\times$1}\nl
\multicolumn{4}{r}{Image Area $A$:} & \multicolumn{3}{l}{22.6 arcsec$^{2}$}

\nl
\end{planotable}

\clearpage

\begin{planotable}{crrrrrr}
\tablewidth{265pt}
\tablecaption{NGC 2808 Star Counts.}
\tablehead{
\colhead{$r$} &
\colhead{$r_{eff}$} &
\colhead{N} &
\colhead{f$_{bkd}$} &
\colhead{$\alpha_c$} &
\colhead{f} &
\colhead{Error}}

\startdata

 0 -   1&  0.71&     0&  3.766&\nodata&\nodata&\nodata\nl
 1 -   2&  1.58&     9&  3.766&\nodata&\nodata&\nodata\nl
 2 -   3&  2.55&    86&  3.766&\nodata&\nodata&\nodata\nl
 3 -   4&  3.54&   199&  3.766& 6.246&  52.755&   4.007\nl
 4 -   5&  4.53&   271&  3.766& 2.574&  20.906&   1.499\nl
 5 -   6&  5.52&   286&  3.765& 1.806&  11.184&   0.884\nl
 6 -   7&  6.52&   272&  3.765& 1.522&   6.375&   0.615\nl
 7 -   8&  7.52&   234&  3.765& 1.354&   2.957&   0.439\nl
 8 -   9&  8.51&   223&  3.764& 1.305&   1.685&   0.365\nl
 9 -  10&  9.51&   205&  3.764& 1.264&   0.577&   0.303\nl
10 -  12& 11.05&   482&  3.763& 1.271&   0.668&   0.202\nl
12 -  14& 13.04&   489&  3.762& 1.267&   0.032&   0.172\nl
14 -  16& 15.03&   557&  3.761& 1.235&  -0.110&   0.155\nl
16 -  18& 17.03&   671&  3.759& 1.251&   0.171&   0.152\nl
18 -  20& 19.03&   666&  3.757& 1.240&  -0.297&   0.134\nl
20 -  25& 22.64&  2154&  3.754& 1.250&   0.057&   0.082\nl
25 -  30& 27.61&  2455&  3.747& 1.238&  -0.229&   0.071\nl
30 -  35& 32.60&  3101&  3.740& 1.248&   0.050&   0.068\nl
35 -  40& 37.58&  3457&  3.731& 1.243&  -0.083&   0.062\nl
40 -  50& 45.28&  8442&  3.716& 1.245&   0.001&   0.040\nl
50 -  60& 55.23& 10208&  3.692& 1.249&  -0.003&   0.037\nl
60 -  70& 65.19& 11974&  3.664& 1.250&   0.001&   0.033\nl
70 -  80& 75.17& 13699&  3.632& 1.248&  -0.005&   0.031\nl
80 -  90& 85.15& 15746&  3.596& 1.248&   0.084&   0.029\nl
90 - 100& 95.13& 17165&  3.557& 1.239&   0.005&   0.027\nl

\nl
\multicolumn{4}{r}{Sectors used:} & \multicolumn{3}{l}{1-8}\nl
\multicolumn{4}{r}{Surface Fit Order:} &
\multicolumn{3}{l}{2$\times$2}\nl
\multicolumn{4}{r}{Image Area $A$:} & \multicolumn{3}{l}{47.7 arcsec$^{2}$}
\nl

\end{planotable}

\clearpage

\begin{planotable}{crrrrrr}
\tablewidth{265pt}
\tablecaption{NGC 3201 Star Counts.}
\tablehead{
\colhead{$r$} &
\colhead{$r_{eff}$} &
\colhead{N} &
\colhead{f$_{bkd}$} &
\colhead{$\alpha_c$} &
\colhead{f} &
\colhead{Error}}

\startdata

 0 -   1&  0.71&    14&  5.937&\nodata&\nodata&\nodata\nl
 1 -   2&  1.58&   147&  5.937&\nodata&\nodata&\nodata\nl
 2 -   3&  2.55&   388&  5.937&\nodata&\nodata&\nodata\nl
 3 -   4&  3.54&   594&  5.938&\nodata&\nodata&\nodata\nl
 4 -   5&  4.53&   757&  5.938& 3.090&  76.779&   3.006\nl
 5 -   6&  5.52&   859&  5.939& 2.017&  44.195&   1.711\nl
 6 -   7&  6.52&   852&  5.940& 1.679&  29.077&   1.200\nl
 7 -   8&  7.52&   786&  5.940& 1.443&  18.120&   0.858\nl
 8 -   9&  8.51&   770&  5.942& 1.386&  14.044&   0.720\nl
 9 -  10&  9.51&   753&  5.943& 1.305&  10.516&   0.600\nl
10 -  12& 11.05&  1379&  5.945& 1.240&   6.422&   0.333\nl
12 -  14& 13.04&  1313&  5.948& 1.199&   3.692&   0.266\nl
14 -  16& 15.03&  1336&  5.952& 1.180&   2.408&   0.229\nl
16 -  18& 17.03&  1425&  5.956& 1.170&   1.845&   0.207\nl
18 -  20& 19.03&  1397&  5.961& 1.152&   0.779&   0.180\nl
20 -  25& 22.64&  4125&  5.971& 1.153&   0.758&   0.105\nl
25 -  30& 27.61&  4793&  5.988& 1.145&   0.365&   0.092\nl
30 -  35& 32.60&  5393&  6.008& 1.138&   0.005&   0.082\nl
35 -  40& 37.58&  6351&  6.031& 1.142&   0.124&   0.077\nl
40 -  50& 45.28& 14599&  6.071& 1.138&  -0.196&   0.049\nl
50 -  60& 55.23& 18438&  6.135& 1.141&  -0.044&   0.045\nl
60 -  70& 65.19& 22059&  6.207& 1.143&  -0.036&   0.042\nl
70 -  80& 75.17& 25859&  6.289& 1.147&   0.004&   0.039\nl
80 -  90& 85.15& 29678&  6.376& 1.148&   0.005&   0.037\nl
90 - 100& 95.13& 33793&  6.468& 1.151&   0.046&   0.035\nl

\nl
\multicolumn{4}{r}{Sectors used:} & \multicolumn{3}{l}{1-8}\nl
\multicolumn{4}{r}{Surface Fit Order:} &
\multicolumn{3}{l}{2$\times$2}\nl
\multicolumn{4}{r}{Image Area $A$:} & \multicolumn{3}{l}{24.1 arcsec$^{2}$}

\nl
\end{planotable}

\clearpage

\begin{planotable}{crrrrrr}
\tablewidth{265pt}
\tablecaption{NGC 4590 Star Counts.}
\tablehead{
\colhead{$r$} &
\colhead{$r_{eff}$} &
\colhead{N} &
\colhead{f$_{bkd}$} &
\colhead{$\alpha_c$} &
\colhead{f} &
\colhead{Error}}

\startdata

 0 -   1&  0.71&     2&  0.263&\nodata&\nodata&\nodata\nl
 1 -   2&  1.58&    82&  0.263&\nodata&\nodata&\nodata\nl
 2 -   3&  2.55&   232&  0.263& 2.261&  33.131&   2.192\nl
 3 -   4&  3.54&   272&  0.263& 1.524&  18.584&   1.143\nl
 4 -   5&  4.53&   250&  0.263& 1.286&  11.111&   0.719\nl
 5 -   6&  5.52&   239&  0.263& 1.192&   7.984&   0.533\nl
 6 -   7&  6.52&   167&  0.263& 1.121&   4.322&   0.355\nl
 7 -   8&  7.52&   130&  0.263& 1.092&   2.750&   0.264\nl
 8 -   9&  8.51&   118&  0.263& 1.070&   2.101&   0.218\nl
 9 -  10&  9.51&    88&  0.263& 1.064&   1.306&   0.167\nl
10 -  12& 11.05&   144&  0.263& 1.051&   0.832&   0.091\nl
12 -  14& 13.04&   127&  0.263& 1.041&   0.547&   0.072\nl
14 -  16& 15.03&   104&  0.263& 1.037&   0.310&   0.056\nl
16 -  18& 17.03&    85&  0.263& 1.034&   0.149&   0.045\nl
18 -  20& 19.03&    90&  0.263& 1.035&   0.127&   0.041\nl
20 -  25& 22.64&   196&  0.263& 1.032&   0.024&   0.020\nl
25 -  30& 27.61&   245&  0.263& 1.033&   0.030&   0.019\nl
30 -  35& 32.60&   261&  0.263& 1.032&   0.001&   0.016\nl
35 -  40& 37.58&   348&  0.263& 1.032&   0.042&   0.016\nl
40 -  50& 45.28&   694&  0.263& 1.031&  -0.010&   0.010\nl
50 -  60& 55.23&   886&  0.263& 1.033&   0.002&   0.009\nl
60 -  70& 65.19&  1050&  0.263& 1.032&   0.003&   0.008\nl
70 -  80& 75.17&  1121&  0.263& 1.032&  -0.017&   0.007\nl
80 -  90& 85.15&  1385&  0.263& 1.033&   0.005&   0.007\nl
90 - 100& 95.13&  1546&  0.263& 1.033&   0.005&   0.007\nl

\nl
\multicolumn{4}{r}{Sectors used:} & \multicolumn{3}{l}{1-8}\nl
\multicolumn{4}{r}{Surface Fit Order:} &
\multicolumn{3}{l}{1$\times$1}\nl
\multicolumn{4}{r}{Image Area $A$:} & \multicolumn{3}{l}{28.9 arcsec$^{2}$}
\nl

\end{planotable}

\clearpage

\begin{planotable}{crrrrrr}
\tablewidth{265pt}
\tablecaption{NGC 5824 Star Counts.}
\tablehead{
\colhead{$r$} &
\colhead{$r_{eff}$} &
\colhead{N} &
\colhead{f$_{bkd}$} &
\colhead{$\alpha_c$} &
\colhead{f} &
\colhead{Error}}

\startdata

 0 -   1&  0.71&    23&  2.719&\nodata&\nodata&\nodata\nl
 1 -   2&  1.58&   158&  2.719& 2.551&  40.049&   3.402\nl
 2 -   3&  2.55&   172&  2.719& 1.412&  12.740&   1.179\nl
 3 -   4&  3.54&   147&  2.719& 1.213&   5.387&   0.669\nl
 4 -   5&  4.53&   132&  2.719& 1.152&   2.659&   0.468\nl
 5 -   6&  5.52&   126&  2.719& 1.145&   1.457&   0.372\nl
 6 -   7&  6.52&   145&  2.719& 1.145&   1.347&   0.338\nl
 7 -   8&  7.52&   170&  2.719& 1.140&   1.394&   0.315\nl
 8 -   9&  8.51&   155&  2.719& 1.123&   0.542&   0.262\nl
 9 -  10&  9.51&   162&  2.719& 1.109&   0.291&   0.236\nl
10 -  12& 11.05&   394&  2.719& 1.114&   0.456&   0.160\nl
12 -  14& 13.04&   433&  2.719& 1.115&   0.238&   0.142\nl
14 -  16& 15.03&   503&  2.719& 1.114&   0.255&   0.133\nl
16 -  18& 17.03&   564&  2.719& 1.111&   0.216&   0.124\nl
18 -  20& 19.03&   626&  2.719& 1.107&   0.185&   0.116\nl
20 -  25& 22.64&  1762&  2.719& 1.107&   0.041&   0.066\nl
25 -  30& 27.61&  2166&  2.719& 1.105&   0.052&   0.060\nl
30 -  35& 32.60&  2533&  2.719& 1.107&   0.028&   0.055\nl
35 -  40& 37.58&  2954&  2.719& 1.111&   0.067&   0.051\nl
40 -  50& 45.28&  7041&  2.719& 1.109&   0.043&   0.033\nl
50 -  60& 55.23&  8440&  2.719& 1.110&  -0.008&   0.030\nl
60 -  70& 65.19& 10153&  2.719& 1.111&   0.043&   0.027\nl
70 -  80& 75.17& 11568&  2.719& 1.109&   0.003&   0.025\nl
80 -  90& 85.15& 13211&  2.719& 1.109&   0.026&   0.024\nl
90 - 100& 95.13& 14536&  2.719& 1.109&  -0.017&   0.022\nl

\nl
\multicolumn{4}{r}{Sectors used:} & \multicolumn{3}{l}{1-8}\nl
\multicolumn{4}{r}{Surface Fit Order:} &
\multicolumn{3}{l}{1$\times$1}\nl
\multicolumn{4}{r}{Image Area $A$:} & \multicolumn{3}{l}{27.2 arcsec$^{2}$}
\nl

\nl
\end{planotable}

\clearpage

\begin{planotable}{crrrrrr}
\tablewidth{265pt}
\tablecaption{NGC 6864 Star Counts.}
\tablehead{
\colhead{$r$} &
\colhead{$r_{eff}$} &
\colhead{N} &
\colhead{f$_{bkd}$} &
\colhead{$\alpha_c$} &
\colhead{f} &
\colhead{Error}}

\startdata

 0 -   1&  0.71&     8&  1.307&\nodata&\nodata&\nodata\nl
 1 -   2&  1.58&    84&  1.307& 2.126&  20.345&   2.362\nl
 2 -   3&  2.55&    83&  1.306& 1.326&   6.703&   0.879\nl
 3 -   4&  3.54&    58&  1.306& 1.148&   2.154&   0.454\nl
 4 -   5&  4.53&    48&  1.306& 1.118&   0.863&   0.313\nl
 5 -   6&  5.52&    44&  1.305& 1.088&   0.277&   0.239\nl
 6 -   7&  6.52&    56&  1.305& 1.095&   0.412&   0.229\nl
 7 -   8&  7.52&    53&  1.304& 1.087&   0.093&   0.192\nl
 8 -   9&  8.51&    62&  1.304& 1.091&   0.144&   0.184\nl
 9 -  10&  9.51&    54&  1.304& 1.080&  -0.187&   0.152\nl
10 -  12& 11.05&   162&  1.303& 1.098&   0.168&   0.116\nl
12 -  14& 13.04&   207&  1.302& 1.091&   0.277&   0.110\nl
14 -  16& 15.03&   215&  1.301& 1.089&   0.118&   0.097\nl
16 -  18& 17.03&   220&  1.301& 1.091&  -0.016&   0.087\nl
18 -  20& 19.03&   266&  1.300& 1.094&   0.093&   0.085\nl
20 -  25& 22.64&   747&  1.298& 1.088&   0.015&   0.048\nl
25 -  30& 27.61&   955&  1.297& 1.092&   0.084&   0.045\nl
30 -  35& 32.60&  1078&  1.295& 1.088&   0.018&   0.040\nl
35 -  40& 37.58&  1257&  1.293& 1.089&   0.036&   0.037\nl
40 -  50& 45.28&  3016&  1.290& 1.088&   0.036&   0.024\nl
50 -  60& 55.23&  3608&  1.286& 1.089&   0.013&   0.022\nl
60 -  70& 65.19&  4243&  1.282& 1.087&   0.009&   0.020\nl
70 -  80& 75.17&  4924&  1.278& 1.090&   0.023&   0.019\nl
80 -  90& 85.15&  5432&  1.274& 1.089&  -0.008&   0.017\nl
90 - 100& 95.13&  6064&  1.270& 1.090&  -0.005&   0.016\nl

\nl
\multicolumn{4}{r}{Sectors used:} & \multicolumn{3}{l}{1-3,7-8}\nl
\multicolumn{4}{r}{Surface Fit Order:} &
\multicolumn{3}{l}{1$\times$1}\nl
\multicolumn{4}{r}{Image Area $A$:} & \multicolumn{3}{l}{37.8 arcsec$^{2}$}
\nl

\end{planotable}

\clearpage

\begin{planotable}{crrrrrr}
\tablewidth{265pt}
\tablecaption{NGC 6934 Star Counts.}
\tablehead{
\colhead{$r$} &
\colhead{$r_{eff}$} &
\colhead{N} &
\colhead{f$_{bkd}$} &
\colhead{$\alpha_c$} &
\colhead{f} &
\colhead{Error}}

\startdata

 0 -   1&  0.71&     4&  1.019&\nodata&\nodata&\nodata\nl
 1 -   2&  1.58&    65&  1.019&\nodata&\nodata&\nodata\nl
 2 -   3&  2.55&   148&  1.019& 2.076&  18.537&   1.608\nl
 3 -   4&  3.54&   111&  1.019& 1.409&   6.091&   0.675\nl
 4 -   5&  4.53&    85&  1.019& 1.240&   2.709&   0.404\nl
 5 -   6&  5.52&    53&  1.019& 1.195&   0.813&   0.252\nl
 6 -   7&  6.52&    57&  1.019& 1.199&   0.655&   0.222\nl
 7 -   8&  7.52&    65&  1.019& 1.189&   0.621&   0.203\nl
 8 -   9&  8.51&    59&  1.019& 1.155&   0.257&   0.166\nl
 9 -  10&  9.51&    52&  1.019& 1.154&  -0.014&   0.139\nl
10 -  12& 11.05&   123&  1.019& 1.172&   0.023&   0.094\nl
12 -  14& 13.04&   164&  1.019& 1.179&   0.164&   0.092\nl
14 -  16& 15.03&   181&  1.019& 1.168&   0.103&   0.083\nl
16 -  18& 17.03&   207&  1.019& 1.178&   0.122&   0.079\nl
18 -  20& 19.03&   183&  1.019& 1.165&  -0.126&   0.066\nl
20 -  25& 22.64&   635&  1.019& 1.170&   0.032&   0.042\nl
25 -  30& 27.61&   786&  1.019& 1.173&   0.048&   0.038\nl
30 -  35& 32.60&   861&  1.019& 1.167&  -0.035&   0.034\nl
35 -  40& 37.58&  1023&  1.019& 1.170&  -0.004&   0.032\nl
40 -  50& 45.28&  2510&  1.019& 1.169&   0.018&   0.021\nl
50 -  60& 55.23&  3208&  1.019& 1.170&   0.067&   0.019\nl
60 -  70& 65.19&  3680&  1.019& 1.168&   0.033&   0.017\nl
70 -  80& 75.17&  4066&  1.019& 1.166&  -0.013&   0.016\nl
80 -  90& 85.15&  4455&  1.019& 1.168&  -0.045&   0.015\nl
90 - 100& 95.13&  5051&  1.019& 1.165&  -0.034&   0.014\nl

\nl
\multicolumn{4}{r}{Sectors used:} & \multicolumn{3}{l}{1-8}\nl
\multicolumn{4}{r}{Surface Fit Order:} &
\multicolumn{3}{l}{1$\times$1}\nl
\multicolumn{4}{r}{Image Area $A$:} & \multicolumn{3}{l}{35.8 arcsec$^{2}$}
\nl

\end{planotable}

\clearpage

\begin{planotable}{crrrrrr}
\tablewidth{265pt}
\tablecaption{NGC 6981 Star Counts.}
\tablehead{
\colhead{$r$} &
\colhead{$r_{eff}$} &
\colhead{N} &
\colhead{f$_{bkd}$} &
\colhead{$\alpha_c$} &
\colhead{f} &
\colhead{Error}}

\startdata

 0 -   1&  0.71&     4&  0.782&\nodata&\nodata&\nodata\nl
 1 -   2&  1.58&   141&  0.782&\nodata&\nodata&\nodata\nl
 2 -   3&  2.55&   244&  0.782& 1.874&  28.324&   1.863\nl
 3 -   4&  3.54&   158&  0.782& 1.220&   7.980&   0.697\nl
 4 -   5&  4.53&    76&  0.782& 1.109&   2.200&   0.342\nl
 5 -   6&  5.52&    57&  0.782& 1.076&   0.993&   0.235\nl
 6 -   7&  6.52&    59&  0.782& 1.079&   0.776&   0.203\nl
 7 -   8&  7.52&    46&  0.782& 1.064&   0.257&   0.153\nl
 8 -   9&  8.51&    53&  0.782& 1.073&   0.283&   0.146\nl
 9 -  10&  9.51&    43&  0.782& 1.064&  -0.016&   0.117\nl
10 -  12& 11.05&   105&  0.782& 1.060&   0.023&   0.079\nl
12 -  14& 13.04&   107&  0.782& 1.063&  -0.086&   0.067\nl
14 -  16& 15.03&   142&  0.782& 1.066&   0.021&   0.067\nl
16 -  18& 17.03&   145&  0.782& 1.064&  -0.060&   0.060\nl
18 -  20& 19.03&   173&  0.782& 1.064&  -0.011&   0.059\nl
20 -  25& 22.64&   499&  0.782& 1.064&  -0.031&   0.034\nl
25 -  30& 27.61&   617&  0.782& 1.063&  -0.023&   0.031\nl
30 -  35& 32.60&   703&  0.782& 1.063&  -0.050&   0.028\nl
35 -  40& 37.58&   875&  0.782& 1.064&   0.008&   0.027\nl
40 -  50& 45.28&  2086&  0.782& 1.067&   0.005&   0.017\nl
50 -  60& 55.23&  2589&  0.782& 1.068&   0.018&   0.016\nl
60 -  70& 65.19&  3025&  0.782& 1.068&   0.009&   0.014\nl
70 -  80& 75.17&  3480&  0.782& 1.068&   0.007&   0.013\nl
80 -  90& 85.15&  3983&  0.782& 1.069&   0.015&   0.013\nl
90 - 100& 95.13&  4305&  0.782& 1.068&  -0.012&   0.012\nl

\nl
\multicolumn{4}{r}{Sectors used:} & \multicolumn{3}{l}{1-8}\nl
\multicolumn{4}{r}{Surface Fit Order:} &
\multicolumn{3}{l}{1$\times$1}\nl
\multicolumn{4}{r}{Image Area $A$:} & \multicolumn{3}{l}{25.7 arcsec$^{2}$}
\nl

\end{planotable}

\clearpage

\begin{planotable}{crrrrrr}
\tablewidth{265pt}
\tablecaption{NGC 7078 Star Counts.}
\tablehead{
\colhead{$r$} &
\colhead{$r_{eff}$} &
\colhead{N} &
\colhead{f$_{bkd}$} &
\colhead{$\alpha_c$} &
\colhead{f} &
\colhead{Error}}

\startdata

 0 -   1&  0.71&     0&  0.236&\nodata&\nodata&\nodata\nl
 1 -   2&  1.58&    13&  0.236&\nodata&\nodata&\nodata\nl
 2 -   3&  2.55&    87&  0.236&\nodata&\nodata&\nodata\nl
 3 -   4&  3.54&   248&  0.236&\nodata&\nodata&\nodata\nl
 4 -   5&  4.53&   404&  0.236&\nodata&\nodata&\nodata\nl
 5 -   6&  5.52&   425&  0.236& 2.012&  24.505&   1.200\nl
 6 -   7&  6.52&   378&  0.236& 1.563&  14.232&   0.744\nl
 7 -   8&  7.52&   319&  0.236& 1.323&   8.716&   0.501\nl
 8 -   9&  8.51&   260&  0.236& 1.230&   5.751&   0.371\nl
 9 -  10&  9.51&   206&  0.236& 1.197&   3.895&   0.288\nl
10 -  12& 11.05&   310&  0.236& 1.129&   2.295&   0.144\nl
12 -  14& 13.04&   200&  0.236& 1.097&   1.106&   0.095\nl
14 -  16& 15.03&   123&  0.236& 1.077&   0.467&   0.063\nl
16 -  18& 17.03&    89&  0.236& 1.075&   0.211&   0.047\nl
18 -  20& 19.03&    87&  0.236& 1.072&   0.154&   0.042\nl
20 -  25& 22.64&   194&  0.236& 1.068&   0.057&   0.021\nl
25 -  30& 27.61&   204&  0.236& 1.067&   0.015&   0.018\nl
30 -  35& 32.60&   210&  0.236& 1.069&  -0.017&   0.015\nl
35 -  40& 37.58&   228&  0.236& 1.066&  -0.030&   0.014\nl
40 -  50& 45.28&   588&  0.236& 1.067&  -0.015&   0.009\nl
50 -  60& 55.23&   756&  0.236& 1.068&  -0.003&   0.008\nl
60 -  70& 65.19&   897&  0.236& 1.068&  -0.002&   0.008\nl
70 -  80& 75.17&   995&  0.236& 1.067&  -0.011&   0.007\nl
80 -  90& 85.15&  1177&  0.236& 1.067&  -0.001&   0.007\nl
90 - 100& 95.13&  1368&  0.236& 1.068&   0.008&   0.007\nl

\nl
\multicolumn{4}{r}{Sectors used:} & \multicolumn{3}{l}{1-8}\nl
\multicolumn{4}{r}{Surface Fit Order:} &
\multicolumn{3}{l}{1$\times$1}\nl
\multicolumn{4}{r}{Image Area $A$:} & \multicolumn{3}{l}{37.8 arcsec$^{2}$}
\nl

\end{planotable}

\clearpage

\begin{planotable}{crrrrrr}
\tablewidth{265pt}
\tablecaption{NGC 7089 Star Counts.}
\tablehead{
\colhead{$r$} &
\colhead{$r_{eff}$} &
\colhead{N} &
\colhead{f$_{bkd}$} &
\colhead{$\alpha_c$} &
\colhead{f} &
\colhead{Error}}

\startdata

 0 -   1&  0.71&     0&  0.306&\nodata&\nodata&\nodata\nl
 1 -   2&  1.58&    16&  0.306&\nodata&\nodata&\nodata\nl
 2 -   3&  2.55&   138&  0.306&\nodata&\nodata&\nodata\nl
 3 -   4&  3.54&   336&  0.306&\nodata&\nodata&\nodata\nl
 4 -   5&  4.53&   484&  0.306&\nodata&\nodata&\nodata\nl
 5 -   6&  5.52&   441&  0.306& 1.960&  24.702&   1.191\nl
 6 -   7&  6.52&   363&  0.306& 1.396&  12.106&   0.651\nl
 7 -   8&  7.52&   262&  0.306& 1.223&   6.493&   0.420\nl
 8 -   9&  8.51&   165&  0.306& 1.146&   3.234&   0.276\nl
 9 -  10&  9.51&   114&  0.306& 1.107&   1.809&   0.198\nl
10 -  12& 11.05&   147&  0.306& 1.081&   0.844&   0.095\nl
12 -  14& 13.04&   113&  0.306& 1.066&   0.431&   0.069\nl
14 -  16& 15.03&    83&  0.306& 1.062&   0.162&   0.051\nl
16 -  18& 17.03&    83&  0.306& 1.056&   0.104&   0.045\nl
18 -  20& 19.03&    96&  0.306& 1.061&   0.120&   0.044\nl
20 -  25& 22.64&   248&  0.306& 1.057&   0.065&   0.024\nl
25 -  30& 27.61&   279&  0.306& 1.058&   0.036&   0.020\nl
30 -  35& 32.60&   317&  0.306& 1.059&   0.023&   0.018\nl
35 -  40& 37.58&   366&  0.306& 1.058&   0.023&   0.017\nl
40 -  50& 45.28&   840&  0.306& 1.057&   0.008&   0.011\nl
50 -  60& 55.23&   986&  0.306& 1.058&  -0.004&   0.010\nl
60 -  70& 65.19&  1218&  0.306& 1.059&   0.010&   0.009\nl
70 -  80& 75.17&  1402&  0.306& 1.059&   0.009&   0.008\nl
80 -  90& 85.15&  1520&  0.306& 1.058&  -0.005&   0.008\nl
90 - 100& 95.13&  1699&  0.306& 1.056&  -0.005&   0.007\nl
\nl
\multicolumn{4}{r}{Sectors used:} & \multicolumn{3}{l}{1-8}\nl
\multicolumn{4}{r}{Surface Fit Order:} &
\multicolumn{3}{l}{1$\times$1}\nl
\multicolumn{4}{r}{Image Area $A$:} & \multicolumn{3}{l}{34.0 arcsec$^{2}$}
\nl

\end{planotable}

\clearpage

\begin{planotable}{ll}
\tablecaption{Sources of Surface Density Data.}
\tablewidth{0pt}
\tablehead{
\colhead{Code} &
\colhead{Source}}

\startdata

CF  & Chun \& Freeman 1979\nl
CJP & Peterson 1986b\nl
CP  & Peterson 1976\nl
GFIQ & Present Study\nl
HAM & Hamuy 1984\nl
HB  & Hanes \& Brodie 1985\nl
HLJ & Johnson 1959\nl
HST & Lauer \etal 1991\nl
II & Illingworth \& Illingworth 1976\nl
KHHW & King \etal 1968\nl
KM  &  Kron \& Mayall 1960\nl
KR  &  Kron \& Gordon 1986\nl

\label{tab:sources}
\end{planotable}

\clearpage

\begin{planotable}{lrrlrcr}
\tablecaption{Fitted Tidal Radii.}
\tablewidth{15cm}
\tablehead{
\colhead{Cluster} &
\colhead{ $r_c$} &
\colhead{$r_t$} &
\colhead{Fitted Region} &
\colhead{$\tilde{\chi}^2$} &
\colhead{90\% Confidence Interval} &
\nl
\colhead{} &
\colhead{(arcmin)} &
\colhead{(arcmin)} &
\colhead{(arcmin)} &
\colhead{} &
\colhead{(arcmin)} &}

\startdata

NGC 288 & 1.40  & 17.4 & 4 - 12 & 0.43 & 16.7 - 17.8 & \nl
NGC 362 & 0.20  & 14.2 & 4 - 10 & 4.32 & 13.9 - 14.5 & \nl
NGC 1904 & 0.23 & 11.0 & 2 - 7  & 1.80 & 10.8 - 11.7 & \nl
NGC 2808 & 0.28 & 14.5 & 4 - 15 & 1.20 & 14.0 - 15.0 & \nl
NGC 3201 & 1.10 & 34.7 & 4 - 20 & 3.08 & 33.2 - 36.2 & \nl
NGC 4590 & 0.80 & 25.2 & 2 - 12 & 1.31 & 23.2 - 27.0 & \nl
NGC 5824 & 0.01 & 11.6 & 1 - 6  & 0.66 & 10.4 - 13.4 & \nl
NGC 6864 & 0.11 & 9.3  & 1 - 6 & 0.08 &  ~7.7 - 10.9 & \nl
NGC 6934 & 0.22 & 9.7  & 2 - 6 & 0.48 &  ~8.6 - 10.7 & \nl
NGC 6981 & 0.42 & 8.3  & 2 - 6 & 1.68 &  ~7.6 - 8.6~ & \nl
NGC 7078 & 0.22 & 23.2 & 5 - 16& 1.92 &  22.8 - 23.7 & \nl
NGC 7089 & 0.37 & 15.9 & 5 - 12 & 2.88 & 15.7 - 16.3 & \nl

\label{tab:fits}
\end{planotable}